\documentclass[12pt]{article}
\parskip \medskipamount

\usepackage{soul,color}
\soulregister\cite7
\soulregister\ref7
\soulregister\pageref7
\soulregister\citet7
\soulregister\citep7

\newcommand\citePkg[1]{\texttt{#1} \citep{#1}}

\DeclareMathAlphabet{\mathcal}{OMS}{cmsy}{m}{n}

\usepackage{setspace}
\usepackage[a4paper, left=1in, right=1in, top=1.8cm, bottom=1.8cm, includehead, includefoot]{geometry} 	

\usepackage{titlesec}
\titleformat*{\section}{\bfseries}
\titleformat*{\subsection}{\itshape\bfseries}
\titleformat*{\subsubsection}{\itshape}

\usepackage{natbib}
\usepackage{har2nat}
\setcitestyle{authoryear,open={(},close={)}}

\usepackage{amsfonts}

\usepackage{graphicx} 									
\usepackage[small,bf,hang]{caption}			
\usepackage[labelformat=simple]{subcaption}									

\usepackage{wrapfig}										
\usepackage{multirow}
\usepackage[table]{xcolor}										
\definecolor{lightblue}{rgb}{0.13, 0.67, 0.8}
\usepackage[latin1]{inputenc}						
\usepackage{amsmath}										
\allowdisplaybreaks[1]									
\usepackage{amssymb}										
\usepackage{amsthm}	
\usepackage{leftidx}										
\usepackage{booktabs}										
\usepackage{url}                        
\usepackage{eurosym}                    
\usepackage{textcomp}                   
\usepackage{fancyhdr}                   
\usepackage{listings} 									
\usepackage{pdfpages}										
\usepackage{floatrow}
\usepackage[title,titletoc,toc]{appendix}				
\usepackage{enumerate}									
\usepackage[multiple]{footmisc}																		
\usepackage{enumitem}										
\usepackage{mathtools}										
\usepackage{IEEEtrantools}							
\usepackage{authblk}											
\usepackage{tikz}
\usetikzlibrary{snakes}
\usetikzlibrary{shapes,arrows,decorations.pathmorphing,backgrounds,positioning,fit,petri}												
\usepackage[english]{babel}
\usepackage{grffile}
\usepackage{amsfonts}
\usepackage{amssymb}
\usepackage{relsize}
\usepackage{rotating}
\usepackage{siunitx}
\usepackage{color}	
\usepackage{etoolbox}
\usepackage{todonotes}
\usepackage{longtable}
\usepackage[plainpages=false,hyperfootnotes=false]{hyperref}
\hypersetup{
	colorlinks   = true, 
	urlcolor     = blue, 
	linkcolor    = red, 
	citecolor   = blue 
}
\usepackage[capitalize]{cleveref}

\usepackage{bookmark}
\usepackage[ruled,vlined,linesnumbered]{algorithm2e}
\SetAlCapFnt{\small} 
\SetKw{KwTo}{in}

\SetKwInput{kwModel}{Model}
\SetKwInput{kwInit}{Initialization}
\newcommand{\nextnr}{\stepcounter{AlgoLine}\ShowLn}

\usetikzlibrary{tikzmark,fit}
\usetikzlibrary{decorations.pathreplacing,calc}

\makeatletter

\pretocmd{\NAT@citex}{%
	\let\NAT@hyper@\NAT@hyper@citex
	\def\NAT@postnote{#2}%
	\setcounter{NAT@total@cites}{0}%
	\setcounter{NAT@count@cites}{0}%
	\forcsvlist{\stepcounter{NAT@total@cites}\@gobble}{#3}}{}{}
\newcounter{NAT@total@cites}
\newcounter{NAT@count@cites}
\def\NAT@postnote{}

\def\NAT@hyper@citex#1{%
	\stepcounter{NAT@count@cites}%
	\hyper@natlinkstart{\@citeb\@extra@b@citeb}#1%
	\ifnumequal{\value{NAT@count@cites}}{\value{NAT@total@cites}}
	{\ifNAT@swa\else\if*\NAT@postnote*\else%
		\NAT@cmt\NAT@postnote\global\def\NAT@postnote{}\fi\fi}{}%
	\ifNAT@swa\else\if\relax\NAT@date\relax
	\else\NAT@@close\global\let\NAT@nm\@empty\fi\fi
	\hyper@natlinkend}
\renewcommand\hyper@natlinkbreak[2]{#1}

\patchcmd{\NAT@citex}
{\ifNAT@swa\else\if*#2*\else\NAT@cmt#2\fi
	\if\relax\NAT@date\relax\else\NAT@@close\fi\fi}{}{}{}
\patchcmd{\NAT@citex}
{\if\relax\NAT@date\relax\NAT@def@citea\else\NAT@def@citea@close\fi}
{\if\relax\NAT@date\relax\NAT@def@citea\else\NAT@def@citea@space\fi}{}{}

\makeatother

\definecolor{darkgreen}{RGB}{40,150,40}

\newcommand{\bs}{\boldsymbol}

\renewcommand{\epsilon}{\varepsilon}

\theoremstyle{definition}

\theoremstyle{remark}

\newtheoremstyle{assump}
{\topsep}   
{\topsep}   
{\itshape}  
{0pt}       
{\itshape} 
{}         
{5pt plus 1pt minus 1pt} 
{\thmname{#1}\thmnumber{\@ifnotempty{#1}{ }#2}%
	\thmnote{ {\the\thm@notefont(#3)}}}         
\theoremstyle{remark}

\makeatletter
\newtheoremstyle{note}
{3pt}
{3pt}
{}
{}
{\itshape}
{.}
{.5em}
{\thmname{#1}\thmnumber{\@ifnotempty{#1}{ }#2}%
	\thmnote{ {\the\thm@notefont(#3)}}}
\makeatother
\theoremstyle{note}
\newtheorem{test}{Assumption}

\newcommand{\Var}[1]{\mathrm{Var}\! \left[ #1 \right]}

\newcommand{\ignore}[1]{}

\newcolumntype{K}{>{\centering\arraybackslash}m{1.5cm}}
\newcolumntype{L}{>{\centering\arraybackslash}m{2.5cm}}
\newcolumntype{M}{>{\centering\arraybackslash}m{3.5cm}}
\newcolumntype{A}{>{\raggedright\arraybackslash}m{2cm}}
\newcolumntype{B}{>{\raggedright\arraybackslash}m{5cm}}
\newcolumntype{W}{>{\raggedright\arraybackslash}p{12cm}}
\newcolumntype{X}{>{\raggedright\arraybackslash}p{8cm}}
\newcolumntype{N}{>{\raggedright\arraybackslash}p{9.0cm}}
\newcolumntype{C}[1]{>{\centering\let\newline\\\arraybackslash\hspace{0pt}}m{#1}}
\newcolumntype{D}{>{\raggedright\arraybackslash}m{1cm}}
\newcolumntype{E}{>{\raggedright\arraybackslash}m{1.5cm}}
\newcolumntype{H}{>{\setbox0=\hbox\bgroup}c<{\egroup}@{}}
\newcolumntype{I}{>{\centering\arraybackslash}m{2.1cm}}

\usepackage{threeparttable}
\newfloatcommand{capbtabbox}{table}[][\FBwidth]
\newfloatcommand{ImportTable}{input}[][\FBwidth]
\newfloatcommand{capthreepart}{threeparttable}[][\FBwidth]
\DeclareFloatFont{Small}{\fontsize{9}{11}\selectfont}
\DeclareFloatFont{Small2}{\small}
\DeclareFloatFont{footnote}{\footnotesize}
\DeclareFloatFont{scriptsize}{\scriptsize}
\DeclareMarginSet{hangboth}{\setfloatmargins*{\hskip-4cm}{\hskip-4cm}}

\DeclareNewFloatType{figurea}
{placement=htbp, name=Figure, within=section, fileext=lofa}

\newfloatcommand{ffigabox}{figurea}[\nocapbeside][]
\makeatletter
\def\thickhline{%
	\noalign{\ifnum0=`}\fi\hrule \@height \thickarrayrulewidth \futurelet
	\reserved@a\@xthickhline}
\def\@xthickhline{\ifx\reserved@a\thickhline
	\vskip\doublerulesep
	\vskip-\thickarrayrulewidth
	\fi
	\ifnum0=`{\fi}}
\makeatother

\newlength{\thickarrayrulewidth}
\makeatletter
\newcommand{\raisemath}[1]{\mathpalette{\raisem@th{#1}}}
\newcommand{\raisem@th}[3]{\raisebox{#1}{$#2#3$}}
\makeatother

\hypersetup{
	colorlinks = true,
	citecolor  = blue,
	urlcolor   = red,
	breaklinks = true
}

\newcolumntype{O}{>{\centering\arraybackslash}m{1.5cm}}
\renewcommand*{\thesection}{\arabic{section}}
\usepackage{subcaption}

\graphicspath{{Figures/}}

\author[1]{Bavo DC Campo}
\author[1,2,3,4]{Katrien Antonio}
\affil[1]{Faculty of Economics and Business, KU Leuven, Belgium.}
\affil[2]{Faculty of Economics and Business, University of Amsterdam, The Netherlands.}
\affil[3]{LRisk, Leuven Research Center on Insurance and Financial Risk Analysis, KU Leuven, Belgium.}
\affil[4]{LStat, Leuven Statistics Research Center, KU Leuven, Belgium.}
\title{\textbf{Insurance pricing with hierarchically structured data\\ \large An illustration with a workers' compensation insurance portfolio}}
\date{}

\allowdisplaybreaks

\begin{document}
\newcommand{\form}[1]{\scalebox{1.087}{\boldmath{#1}}}

\sloppy
\maketitle

\begin{abstract}
	\noindent
	Actuaries use predictive modeling techniques to assess the loss cost on a contract as a function of observable risk characteristics. State-of-the-art statistical and machine learning methods are not well equipped to handle hierarchically structured risk factors with a large number of levels. In this paper, we demonstrate the data-driven construction of an insurance pricing model when hierarchically structured risk factors, contract-specific as well as externally collected risk factors are available. We examine the pricing of a workers' compensation insurance product with a hierarchical credibility model \citep{JewellModel}, Ohlsson's combination of a generalized linear and a hierarchical credibility model \citep{Ohlsson2008} and mixed models. We compare the predictive performance of these models and evaluate the effect of the distributional assumption on the target variable by comparing linear mixed models with Tweedie generalized linear mixed models. For our case-study the Tweedie distribution is well suited to model and predict the loss cost on a contract. Moreover, incorporating contract-specific risk factors in the model improves the predictive performance and the risk differentiation in our workers' compensation insurance portfolio.\\[5mm]
	\textbf{Keywords:} hierarchically clustered data, Tweedie regression, random effects, hierarchical credibility, generalized linear mixed models, multi-level factor, high-cardinality feature
\end{abstract}

\section{Introduction}
When pricing insurance contracts via risk classification, property and casualty (P\&C or general, non-life) insurers use observable characteristics to group policyholders with a similar risk profile in tariff classes. \textcolor{black}{To construct these tariff classes, we either use supervised or unsupervised learning techniques or a combination of both. For example, \citet{Henckaerts2020} developed a tariff structure using tree-based machine learning methods, \citet{Gao2018} employed clustering techniques to group policyholders with similar driving behavior and \citet{Zhu2021} combined image classification with clustering techniques to differentiate between driving styles.}
	
Actuaries then estimate the loss cost for each constructed tariff class as a function of the observed risk characteristics using supervised learning methods. Within P\&C insurance, \textcolor{black}{continuous and geographical risk factors are typically binned into categorical variables with a limited number of levels. This transformation is either based on expert opinion \citep{Frees2008,Antonio2010} or obtained in a data-driven way \citep{Henckaerts2018}. The categorical format} enables the construction of an interpretable tariff list that is easily explainable to all stakeholders. However, certain types of risk factors pose a challenge when we want to incorporate them in a pricing model. This particularly holds true for hierarchically structured risk factors with a large number of levels, which are also known as high-cardinality risk factors within the machine learning literature \citep{Micci2001} or as multi-level risk factors (MLF) within the actuarial literature \citep{Ohlsson}. In this paper, we illustrate the construction of a data-driven insurance pricing model when both hierarchically structured risk factors and contract-specific risk factors are available.

Currently, generalized linear models (GLMs) \citep{GLM} are regarded as state-of-the-art for insurance pricing \citep{Haberman1996, Dejong2008,Frees2015}. One of the main advantages of GLMs is that the assumed distribution of the response variable belongs to the exponential family, thereby facilitating the modeling of non-normally distributed response variables such as the claim frequency or severity. The frequency-severity decomposition is a popular modeling strategy among P\&C insurers \citep{Denuit2007,Frees2014,Parodi2014, Ohlsson,Henckaerts2018,Henckaerts2020}, where separate predictive models are built for the claim frequency and severity. In this approach we include contracts that reported zero claims during the policy period in the frequency model, but omit these when modeling the claim severity. Alternatively, we can use a Tweedie GLM which enables modeling the zero and continuous positive claim costs simultaneously \citep{Jorgensen1994, SmythTweedie2002, Ohlsson}.\ignore{This approach enables us to incorporate and handle exact zeros in an appropriate manner \citep{Jorgensen1994}.} Recently, the traditional GLM is being challenged by machine learning methods. In contrast to GLMs, such methods are able to learn complex nonlinear transformations and interactions between risk factors without having to specify them explicitly \citep{Blier2021}. \citet{Henckaerts2020} and \citet{Yang2018}, for example, showed how tree-based machine learning methods can be used to develop pricing models that outperform the classical GLM. \textcolor{black}{Notwithstanding, machine learning methods have their own drawbacks. They might be more prone to overfitting \citep{Ying2019,Fang2019,Colbrook2022}, less transparent \citep{MaidR, Dastile2020} and cannot reliably estimate the prediction uncertainty \citep{PredictionUncertainty,Ovadia2019,Tohme2022,Klass2018}.}


Both GLMs and machine learning methods experience difficulties when confronted with MLFs. Within car insurance, a typical example would be the car model. Due to the large number of levels we often have insufficient data to get reliable parameter estimates when using a GLM with car model as a factor variable. Further, machine learning methods become computationally intractable when dummy encoding is applied to the MLFs. Our paper focuses on MLFs that exhibit a hierarchical structure and a typical example hereof, within workers' compensation insurance, is the NACE code. NACE stands for the statistical classification of economic activities in the European community \citep{NACE} and is used as a hierarchical classification system to group similar companies based on their economic activities. The NACE code consists of 4 hierarchical levels. When only using the first two levels, an example of a NACE code would be \textit{A03}. The letter is used to identify the first level and \textit{A} stands for \textit{Agriculture, Forestry and Fishing}. The numbers following the letter identify the second level, nested within the first level. Here, \textit{03} refers to \textit{Fishing and aquaculture}. One way to handle (hierarchical) MLFs is via preprocessing with encoding methods that transform the categorical variable into a continuous one, see e.g. the strategy proposed in \citet{Micci2001}.

Alternatively, we can introduce hierarchically structured random effects into our predictive model to handle the hierarchical MLF. Random effects make optimal use of both the within-cluster and between-cluster claims experience. \textcolor{black}{Applied to our example, at the second level in the hierarchy, the within-cluster claims experience for \textit{A03} refers to the experience obtained from all companies in \textit{03} within \textit{A}. At the first level in the hierarchy, it entails the experience from all companies within cluster $A$. Between-cluster experience refers to the differences observed when comparing the claims experience across different clusters at the first (i.e. clusters \textit{A}, \textit{B}, $\dots$) and second (i.e. clusters within \textit{A}) level in the hierarchy}. Random effects allow to account for within-cluster dependency and between-cluster heterogeneity present in hierarchically structured data and enable the prediction of the loss cost as a function of both the contract-specific risk factors and the hierarchical MLF. \textcolor{black}{Further, to estimate the effect of the hierarchical MLF, we only have to estimate the variance of the (hierarchical) level-specific effects. Consequently, in comparison to dummy encoding, the random effects approach drastically reduces the number of parameters. A drawback of the random effects approach is that their estimation and the interpretation of the model output is more cumbersome than with a traditional GLM \citep{Bolker2009, Zuur2009, Harrison2018}.}

The hierarchical credibility model of \citet{JewellModel} is one of the best-known actuarial random effects models. In this model, only assumptions on the mean and variance of the random variables (i.e. the response variable and the random effects) are made, making it a distribution-free approach. The hierarchical credibility model (or Jewell model, we use these terms interchangeably), however, does not allow the inclusion of contract-specific risk factors. \citet{Ohlsson2008} therefore combined a GLM with the hierarchical credibility model which allows for a distributional assumption on the response. Another approach that makes use of random effects is the mixed models framework. Mixed models extend GLMs to accommodate correlated or clustered responses. In this framework, we impose distributional assumptions on the response, conditional on the random effects, and on the random effects. Within the actuarial literature, there are numerous papers that illustrate and advocate their use in ratemaking. Moreover, \citet{FreesIME} showed how several credibility models, including the hierarchical credibility model, can be expressed as special cases of the linear mixed model (LMM). \citet{AntonioGLMM} gave a detailed overview of the theory and actuarial applications of generalized linear mixed models (GLMMs) as well as several advantages of using GLMMs. Another illustration is given in \citet{Antonio2010}, where a hierarchically structured intercompany claim data set on fleet contracts was analyzed using GLMMs and Bayesian estimation techniques. 

\textcolor{black}{Our paper contributes to the actuarial literature in three ways. First, we provide a detailed discussion (with strengths and weaknesses) of pricing a workers' compensation insurance product with the hierarchical credibility model \cite{JewellModel}, Ohlsson's combination of a generalized linear and a hierarchical credibility model \cite{Ohlsson2008} and via the framework of (generalized and linear) mixed models. Second, we develop and demonstrate a comprehensive, data-driven workflow for the use of continuous and spatial covariates in such pricing models. Third, we compare the predictive performance of these models and evaluate the effect of the distributional assumption on the target variable. Hereto we compare linear mixed models and Tweedie generalized linear mixed models. }

This paper is structured as follows. In \hyperref[sec:2]{Section 2}, we illustrate the general structure of a workers' compensation insurance portfolio and use this as a basis to introduce the theoretical framework. In \hyperref[sec:3]{Section 3}, we present a case study on a workers' compensation insurance product including an exploratory analysis, the pre-processing of the database, the development of predictive models and the evaluation of their predictive performance. We construct models under different distributional assumptions for the outcome variable, using different sets of company-specific risk factors (i.e. no risk factors, internal risk factors only, internal and externally collected risk factors). We assess the effect of the distributional assumption and the added value of internally and externally collected risk factors by comparing the predictive performance of different model specifications. We conclude with a discussion in \hyperref[sec:4]{Section 4}.

\section{Predictive modeling with hierarchically structured data in the presence of observable risk factors}\label{sec:2}

\subsection{Portfolio of hierarchically structured risks}
\begin{figure}[!htbp]
	\centering
	\caption{\label{fig:HierStr}Hierarchical structure of a hypothetical example}
	\begin{tikzpicture}[scale=1.4, snake=zigzag, line before snake = 5mm, line after snake = 5mm]
		
		\draw (6,14) node{\small{$\bullet$}};
		
		\node[anchor = west] at (0, 14.2){Portfolio};
		\node[anchor = west] at (0, 12){Industry};
		\node[anchor = west] at (0, 9.2){Branch};
		\node[anchor = west] at (0, 6.4){Company};
		\node[anchor = west] at (0, 3.6){Year};
		
		\draw[-] (6,14) -- (5 - 2,12);
		\draw[-] (6,14) -- (7 - 2,12);
		\draw[-] (6,14) -- (9 - 1.5,12);
		\draw[-] (6,14) -- (11 - 1.5,12);
		
		\draw (5 - 2,12) node{\tiny{$\bullet$}};
		\draw (7 - 2,12) node{\tiny{$\bullet$}};
		\draw (9 - 1.5,12) node{\tiny{$\bullet$}};
		\draw (11 - 1.5,12) node{\tiny{$\bullet$}};
		
		\draw (5 - 2,12) node[below=3pt]{\color{magenta} Agriculture};
		\draw (7 - 2,12) node[below=3pt]{\color{magenta} Construction};
		\draw (9 - 1.5,12) node[below=3pt]{\color{magenta} Transportation};
		\draw (11 - 1.5,12) node[below=3pt]{\color{magenta} Utilities};
		
		\draw (8 - 1.75,12) node{$\ldots$};

		\draw[-] (7 - 2,11.2)--(6 - 2,9.2);
		\draw[-] (7 - 2,11.2)--(9 - 2,9.2);
		\draw[dashed] (5 - 2,11.2) -- (4.5 - 2,10.5);
		\draw[dashed] (5 - 2,11.2) -- (5 - 2,10.5);
		\draw[dashed] (5 - 2,11.2) -- (5.5 - 2,10.5);
		\draw[dashed] (9 - 1.5,11.2) -- (8.5 - 1.5,10.5);
		\draw[dashed] (9 - 1.5,11.2) -- (9 - 1.5,10.5);
		\draw[dashed] (9 - 1.5,11.2) -- (9.5 - 1.5,10.5);
		\draw[dashed] (11 - 1.5,11.2) -- (10.5 - 1.5,10.5);
		\draw[dashed] (11 - 1.5,11.2) -- (11 - 1.5,10.5);
		\draw[dashed] (11 - 1.5,11.2) -- (11.5 - 1.5,10.5);
		
		\draw (7 - 2,11.2) node{\tiny{$\bullet$}};
		\draw (5 - 2,11.2) node{\tiny{$\bullet$}};
		\draw (9 - 1.5,11.2) node{\tiny{$\bullet$}};
		\draw (11 - 1.5,11.2) node{\tiny{$\bullet$}};

		\draw (6 - 2,9.2) node{\tiny{$\bullet$}};
		\draw (9 - 2,9.2) node{\tiny{$\bullet$}};
		
		\draw (6 - 2,9.2) node[below=3pt]{\color{red} Industrial};
		\draw (9 - 2,9.2) node[below=3pt]{\color{red} Interior design};
		
		\draw (7.5 - 2,9.2) node{$\ldots$};
		
		\draw[-] (6 - 2,8.4)--(5 - 2,6.4);
		\draw[-] (6 - 2,8.4)--(7 - 2,6.4);
		\draw[dashed] (9 - 2,8.4) -- (8.5 - 2,7.4);
		\draw[dashed] (9 - 2,8.4) -- (9 - 2,7.4);
		\draw[dashed] (9 - 2,8.4) -- (9.5 - 2,7.4);
		\draw (9 - 2,8.4) node{\tiny{$\bullet$}};
		
		\draw (6 - 2,8.4) node{\tiny{$\bullet$}};
		\draw (5 - 2,6.4) node{\tiny{$\bullet$}};
		\draw (7 - 2,6.4) node{\tiny{$\bullet$}};
		
		\draw (5 - 2,6.4) node[below=3pt]{$c_1$};
		\draw (7 - 2,6.4) node[below=3pt]{$c_{10}$};
		
		\draw (6 - 2, 6.4) node{$\ldots$};
		\draw (7 - 2, 5.6) node{\tiny{$\bullet$}};
		
		\draw[-] (7 - 2, 5.6)--(4, 3.6);
		\draw[-] (7 - 2, 5.6)--(6, 3.6);
		
		\draw[dashed] (5 - 2, 5.6) -- (4.5 - 2, 4.6);
		\draw[dashed] (5 - 2, 5.6) -- (5 - 2, 4.6);
		\draw[dashed] (5 - 2, 5.6) -- (5.5 - 2, 4.6);
		
		\draw (6 - 2, 3.6) node{\tiny{$\bullet$}};
		\draw (6 - 2, 3.6) node[below=3pt]{$t_1$};
		\draw (8 - 2, 3.6) node{\tiny{$\bullet$}};
		\draw (8 - 2, 3.6) node[below=3pt]{$t_{8}$};
		\draw (7 - 2, 3.6) node{$\ldots$};
	\end{tikzpicture}
\end{figure}
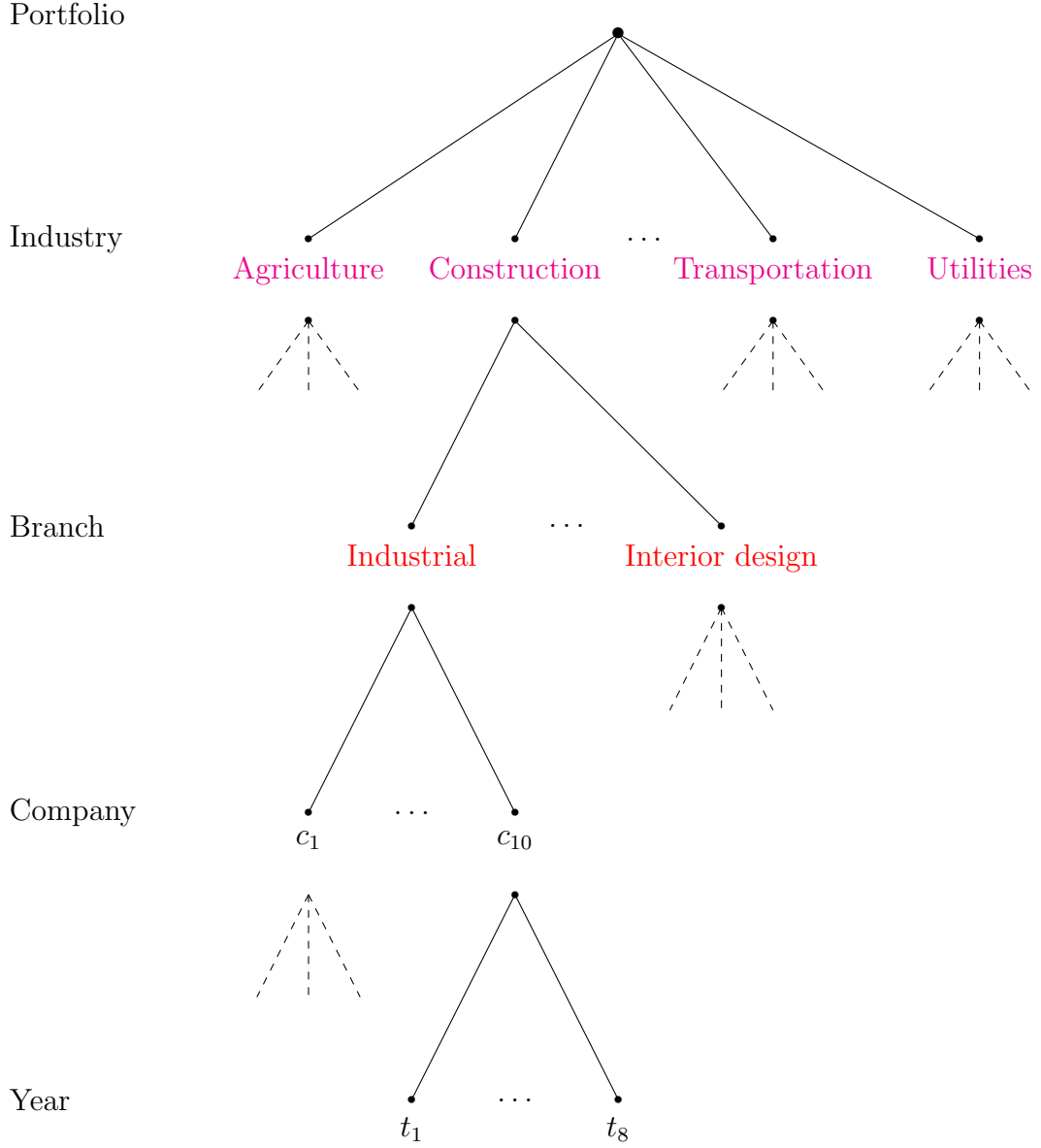

Within insurance pricing, we are interested in determining the loss cost $Y$ defined as the ratio between the claim cost $Z$ and a corresponding exposure measure $w$ of a contract, such as the duration of a policy \citep{Ohlsson}. Some insurance portfolios are characterized by an inherent hierarchical structure and of these, a portfolio of workers' compensation insurance contracts is a prime example. This insurance product provides a financial compensation for lost wages and medical expenses to employees who suffer from a job-related injury \citep{WCdef, Reacfin}. Most often this product is subscribed by the employer, which is typically the company where the employee works. In a workers' compensation insurance product, we commonly define the loss cost $Y$ as the ratio of the total claim amount $Z$ to the salary mass $w$ \citep{RegrModelingActuarial, Denuit2019Book}. To illustrate the typical hierarchical structure of a workers' compensation insurance portfolio, a hypothetical example is given in Figure \ref{fig:HierStr}. We first group the companies into different clusters based on their primary business activity and we refer to this as the industry level. Next, we group the companies via branches within industries. Within each of these branches, we have several companies for which we have yearly data available. Due to the nested structure, there will be heterogeneity between clusters and dependency among observations belonging to the same cluster. It is of utmost importance that this is accurately reflected in our predictive models. 

In addition to the hierarchical MLF, insurance companies use historical data on policies and claims which is their main source of information \citep{Ohlsson} and we refer to this as internal data. Depending on the insurer, information at various hierarchical levels may be available. For example, at the company-level, the insurer may have information on the company size or number of employees. This internally collected data can be supplemented with data from an external source to obtain complementary information on the contracts (e.g. financial statements). This information may help in explaining an additional part of the observed heterogeneity.

Our analysis puts focus on the loss cost $Y_{ijkt}$
\begin{equation}\label{eq:KeyRatio}
	\begin{aligned}
		Y_{ijkt} = \frac{Z_{ijkt}}{w_{ijkt}}
	\end{aligned}
\end{equation}
where we account for inflation by using the year-specific salary mass $w_{ijkt}$. Here, $i$ serves as an index for the risk profile based on the internally and externally collected company-specific risk factors, $j$ denotes the industry, $k$ the branch and $t$ the annual or repeated observations. Using these indices, the company-specific covariate vector is denoted as $\bs{x}_{ijkt}$.

\subsection{Random effects model specification}\label{sec:ModelSpecification}
We specify the following functional form for a random effects model that satisfies our requirements
\begin{equation}\label{eq:GeneralEquation}
	\begin{aligned}
		g(E[Y_{ijkt} | U_j, U_{jk}]) &= \mu + \boldsymbol{x}_{ijkt}^\top \boldsymbol{\beta} + U_j + U_{jk} \\&= \zeta_{ijkt}\\
	\end{aligned}
\end{equation}
\noindent
where $g(\cdot)$ denotes the link function (for example the identity or log link),  $\mu$ the intercept, $\bs{x}_{ijkt}$ the company-specific covariate vector and $\boldsymbol{\beta}$ the corresponding parameter vector. With the model parameters $\mu$ and $\bs{\beta}$ we capture the company-specific effects. To assess the effect of the hierarchical MLF, we introduce the random effects $U_j$ and $U_{jk}$ which capture the unobservable effects of the industry and the branch in which the company operates. $U_j$ denotes the industry-specific deviation from $\mu + \boldsymbol{x}_{ijkt}^\top \boldsymbol{\beta}$ and $U_{jk}$ denotes the branch-specific deviation from $\mu + \boldsymbol{x}_{ijkt}^\top \boldsymbol{\beta} + U_{j}$. We assume that the random industry effects $U_j$ are independent and identically distributed (i.i.d.) with $E[U_j] = 0$ and $\Var{U_j} = \sigma_{I}^2$. Similarly, the random branch effects $U_{jk}$ are assumed to be i.i.d. with $E[U_{jk}] = 0$ and $\Var{U_{jk}} = \sigma_{B}^2$. We do not specify any company-specific random effects since we want to construct an a priori pricing model.

We refer to the right hand-side of equation \eqref{eq:GeneralEquation} as the systematic model component, which specifies how the company-specific covariates and hierarchical MLF are combined with $\mu, \bs{\beta}, U_j$ and $U_{jk}$ to give the linear predictor $\zeta_{ijkt}$. Next to this systematic component, we introduce a distributional assumption for the conditional response $Y_{ijkt} | U_j, U_{jk}$. We assume that the distribution belongs to the exponential family with probability density function (pdf) 
\begin{equation}
	\begin{aligned}
		f(Y_{ijkt} | \mu, \bs{\beta}, U_j, U_{jk}, \phi, w_{ijkt}) = \exp{\left\{ \frac{Y_{ijkt} \theta_{ijkt} - \psi(\theta_{ijkt})}{\phi} w_{ijkt} + c(y_{ijkt}, \phi, w_{ijkt}) \right\} }
	\end{aligned}
\end{equation}
\noindent
where $\psi(\cdot)$ and $c(\cdot)$ are known functions, $\phi$ denotes the dispersion parameter and $\theta_{ijkt}$ is the natural parameter. Further, the following conditional relations hold
\begin{equation}
	\begin{aligned}
		g^{-1}(\zeta_{ijkt}) &= E[Y_{ijkt} | U_j, U_{jk}] = \psi^{'}(\theta_{ijkt}),\\ \Var{Y_{ijkt} | U_{j}, U_{jk}} &= \frac{\phi}{w_{ijkt}} \psi^{''}(\theta_{ijkt}) = \frac{\phi}{w_{ijkt}} V(g^{-1}(\zeta_{ijkt})),
	\end{aligned}
\end{equation}
\noindent
where $V(\cdot)$ denotes the variance function. 

Given the continuous nature of the registered losses, we can choose to model the loss cost by assuming a Gaussian distribution with an identity link function. However, a disadvantage of the Gaussian assumption is the handling of contracts with a zero loss cost (i.e. no claim occurred). Moreover, when fitting a Gaussian distribution on a sample containing zero-valued observations, the resulting fit will implicitly assume the existence of negative values due to the symmetric nature of this distribution. Consequently, this may inadvertently lead to predictions with a negative value which is undesirable in a pricing model. To address this shortcoming of the Gaussian distribution, we can opt for either the frequency-severity or Tweedie approach to appropriately model the zero and non-zero valued observations. The advantage of the Tweedie approach, compared to the frequency-severity strategy, is that we are able to model the claim frequency and severity simultaneously. This allows us to estimate the loss cost directly.

One of the most defining characteristics of the Tweedie distribution (see \citet{Delong2021}, for example) is the relationship between the variance function and the mean
\begin{equation}
	\begin{aligned}
		\Var{Y_{ijkt}| U_j, U_{jk}} = \frac{\phi \cdot (g^{-1}(\zeta_{ijkt}))^p}{w_{ijkt}}
	\end{aligned}
\end{equation}
\noindent
with $p \in (-\infty, 0] \ \cup \ [1, \infty)$. The Tweedie family of distributions encompasses a large range of distributions, which are characterized by the value of $p$ (see Table \ref{tab:PowerParameter}). 

\begin{table}[H]
	\centering
	\caption{\label{tab:PowerParameter}The power parameter $p$ and its associated distribution}
	\begin{tabular}{cc}
		\hline
		Value of p & Distribution\\
		\hline
		$p = 0$ & Normal\\
		$p = 1$ & Poisson\\
		$p \in (1, 2)$ & Compound Poisson - Gamma\\
		$p = 2$ & Gamma\\
		$p = 3$ & Inverse Gaussian\\
		\hline
	\end{tabular}
\end{table}

\subsection{Parameter estimation}
\subsubsection{Hierarchical credibility model}\label{Sec:Jewell}
The basic hierarchical credibility model of \citet{JewellModel} corresponds to a random effects model where no company-specific covariates (i.e. $\boldsymbol{x}_{ijkt}^\top \boldsymbol{\beta} = 0$ in \eqref{eq:GeneralEquation}) are included and where $g(\cdot)$ is the identity link function
\begin{equation}\label{eq:JewellAdditive}
	\begin{aligned}
		E[Y_{ijkt} | U_j, U_{jk}] &= \mu + U_j + U_{jk}.\\
	\end{aligned}
\end{equation}
\noindent
We refer to \eqref{eq:JewellAdditive} as the additive Jewell model. We focus on the estimation of the industry expectation $V_j = \mu + U_j$ and the branch expectation $V_{jk} = \mu + U_j + U_{jk}$ \citep{Dannenburg}. These represent the conditional mean of all observations in industry $j$ and of all observations in branch $k$ within industry $j$, respectively, since $E[Y_{ijkt}|U_j] = V_j$ and $E[Y_{ijkt}|U_j, U_{jk}] = V_{jk}$. Following \citeauthor{OhlssonJewell} \citeyearpar{OhlssonJewell,Ohlsson2008}, we rewrite the hierarchical credibility model of \citet{JewellModel} as
\begin{equation}\label{eq:BasicJewellModel}
	\begin{aligned}
		E[Y_{ijkt} | V_j, V_{jk}] &= V_{jk} \ \ \text{and} \ \ E[Y_{ijkt} | V_j] &= V_j.
	\end{aligned}
\end{equation}
\noindent
We make the following assumptions

\begin{test}\label{ass:JewellAssumptions}
\end{test}
\begin{enumerate}[label=(\alph*), ref={assumption~\Alph*}]
	\item The industries are independent, i.e. ($Y_{ijkt}, V_j, V_{jk}$) and ($Y_{i'j'k't'}, V_{j'}, V_{j'k'}$) are independent for $j \neq j'$.
	\item For every $j$, conditional on the industry effect $V_j$, the branches are independent, i.e. ($Y_{ijkt}, V_{jk}$) and ($Y_{i'jk't'}, V_{jk'}$) are conditionally independent for $k \neq k'$.
	\item All the pairs ($V_j, V_{jk}$), $j = (1, \dots, J)$; $k = 1, \dots, K_j$; are identically distributed, with $E[V_j] = \mu > 0$, $E[V_{jk} | V_j] = V_j$, $\Var{V_j} = \sigma_{I}^2$ and $E[\Var{V_{jk} | V_j}] = \sigma_{B}^2$.
	\item For any ($j$, $k$), conditional on ($V_j, V_{jk}$), the $Y_{ijkt}$ are independent, with mean $V_{jk}$ and with variance satisfying $E[\Var{Y_{ijkt} | V_j, V_{jk}}] = \sigma^2 / w_{ijkt}$.
%
\end{enumerate}
We use $\mu = E[Y_{ijkt}] = E[V_j] = E[V_{jk}]$ to denote the overall expectation. Using Assumption \ref{ass:JewellAssumptions} (c) and (d), it follows that
\begin{equation}
	\begin{aligned}
		\Var{Y_{ijkt}} &= E[\Var{Y_{ijkt}| V_j, V_{jk}}] + \Var{E[Y_{ijkt} | V_j, V_{jk}]}\\
		&= \frac{\sigma^2}{w_{ijkt}} + \sigma_{I}^2 + \sigma_{B}^2.
	\end{aligned}
\end{equation}
\noindent
The credibility estimator of $V_j$ \citep{Dannenburg,OhlssonJewell,Ohlsson2008}, under Assumption \ref{ass:JewellAssumptions}, is defined as
\begin{equation}\label{eq:CredEstVj}
	\begin{aligned}
		\widehat{V}_j &= q_j \bar{Y}_{\cdot j \cdot \cdot}^z + (1 - q_j) \mu,\\
	\end{aligned}
\end{equation}
\noindent
where
\begin{equation}
	\begin{aligned}
		\bar{Y}_{\cdot jk \cdot} &= \frac{\sum_{i, t} w_{ijkt} Y_{ijkt}}{\sum_{i, t} w_{ijkt}},\ \ \ z_{jk} = \frac{w_{\cdot jk \cdot}}{w_{\cdot jk \cdot} + \sigma^2 / \sigma_{B}^2},\\
		\bar{Y}_{\cdot j \cdot \cdot}^z &= \frac{\sum_k z_{jk} \bar{Y}_{\cdot jk \cdot}}{\sum_k z_{jk}}, \ \ \text{and} \ q_j = \frac{z_{j \cdot}}{z_{j \cdot} + \sigma_{B}^2 / \sigma_{I}^2},
	\end{aligned}
\end{equation}
and we define $w_{\cdot jk \cdot} = \sum_{i, t} w_{ijkt}$. The credibility estimator of $V_{jk}$ is specified as
\begin{equation}\label{eq:CredEstVjk}
	\begin{aligned}
		\widehat{V}_{jk} &= z_{jk} \bar{Y}_{\cdot jk \cdot} + (1 - z_{jk}) \widehat{V}_j.\\
	\end{aligned}
\end{equation}
\noindent
Here, $q_j$ and $z_{jk}$ are the credibility factors at the industry- and branch-level, respectively. $\bar{Y}_{\cdot jk \cdot}$ represents the weighted average for the $k^{th}$ branch within industry $j$ and serves as an estimator of the average loss cost at the branch level. The estimator of the average loss cost at the industry level is denoted by $\bar{Y}_{\cdot j \cdot \cdot}^z$ and we use the superscript $z$ to indicate that we weigh the averages $\bar{Y}_{\cdot jk \cdot}$ with the credibility factors $z_{jk}$ instead of the original weights $w_{\cdot jk \cdot}$. The latter estimators, however, are not optimal for clusters that have a low number of observations. We therefore use the credibility estimators $\widehat{V}_j$ and $\widehat{V}_{jk}$, which are a weighted sum of a more stable average and a less stable, more cluster-specific average. To use these credibility estimators, we first require estimators of the variance parameters $\sigma^2, \sigma_{I}^2$ and $\sigma_{B}^2$ as well as an estimator of $\mu$ (see Appendix \ref{App:JewellVariance}). We refer the reader to \citet{Dannenburg}, \citet{OhlssonJewell} and \citet{Ohlsson} for detailed information on these estimators. \textcolor{black}{Next, we predict the damage rate using $\widehat{Y}_{ijkt} = \widehat{V}_{jk}$.}

The hierarchical credibility model is relatively easy to implement and computationally light, which is one of its advantages. Furthermore, we only require estimates of the mean and variance parameters to obtain the random effect estimates. Statistical inference on the estimated parameters, however, is not possible with this distribution-free approach.

\subsubsection{Combining the hierarchical credibility model with a GLM}
 \citet{Ohlsson2008} reformulates the hierarchical credibility model in \eqref{eq:BasicJewellModel} as a multiplicative random effects model by defining $V_j = \tilde{\mu} \ \widetilde{U}_j$ and $V_{jk} = \tilde{\mu} \ \widetilde{U}_j \ \widetilde{U}_{jk} = V_j \widetilde{U}_{jk}$. Consequently,
\begin{equation}\label{eq:JewellMultiplicative}
	\begin{aligned}
		E[Y_{ijkt} | \widetilde{U}_j, \widetilde{U}_{jk}] &= \tilde{\mu} \ \widetilde{U}_j \ \widetilde{U}_{jk}\\
	\end{aligned}
\end{equation}
and we refer to \eqref{eq:JewellMultiplicative} as the multiplicative Jewell model. To obtain this multiplicative structure in \eqref{eq:GeneralEquation} we define $g(\cdot) = \log(\cdot)$. In this case, $\tilde{\mu} = e^{\mu}$, $\widetilde{U}_j = e^{U_j}$ and $\widetilde{U}_{jk} = e^{U_{jk}}$. To allow for company-specific covariates, Ohlsson extends \eqref{eq:JewellMultiplicative} to
\begin{equation}
	\begin{aligned}
		E[Y_{ijkt} | \widetilde{U}_j, \widetilde{U}_{jk}] &= \tilde{\mu} \ \gamma_{ijkt} \ \widetilde{U}_j \ \widetilde{U}_{jk} = \gamma_{ijkt} V_{jk}\\
	\end{aligned}
\end{equation}
\noindent
where $\gamma_{ijkt}$ denotes the effect of the company-specific covariates.  We add a distributional assumption and assume that $Y_{ijkt} | \widetilde{U}_j, \widetilde{U}_{jk} \sim \mathcal{T}(\gamma_{ijkt} V_{jk}, \frac{\phi}{w_{ijkt}} (\gamma_{ijkt} V_{jk})^p))$, where $\mathcal{T}$ denotes any member of the Tweedie family (see \Cref{tab:PowerParameter}). To estimate the parameters in this model, \citet{Ohlsson2008} devised the iterative GLMC (GLMs with credibility) algorithm which is given in Algorithm \ref{algo:GLMC}.

\IncMargin{1em}
\LinesNotNumbered
\begin{algorithm}[!htbp]
	\SetAlgoLined
	\kwModel{$E[Y_{ijkt} | \widetilde{U}_j, \widetilde{U}_{jk}] = \tilde{\mu} \ \gamma_{ijkt} \ \widetilde{U}_j \ \widetilde{U}_{jk}$}
	\kwInit{Set $\widehat{\widetilde{U}}_j = \widehat{\widetilde{U}}_{jk} = 1$}
	\Repeat{convergence}{
		\nextnr
		Estimate $\tilde{\mu}$, $\gamma_{ijkt}$ and $p$ using a GLM with log link, include the $\log(\widehat{\widetilde{U}}_j)$ and $\log(\widehat{\widetilde{U}}_{jk})$ as offset variables and the $w_{ijkt}$'s as weights. This yields $\hat{\tilde{\mu}}$, $\hat{\gamma}_{ijkt}$ and $\hat{p}$\;\label{Step1}
		\nextnr
		Use $\hat{\tilde{\mu}}$ and $\hat{\gamma}_{ijkt}$ to estimate $\sigma^2$, $\sigma^2_{B}$ and $\sigma^2_{I}$ with the hierarchical credibility model \citep{Dannenburg,OhlssonJewell,Ohlsson2008} \; \label{Step2}
		\nextnr
		Compute $\widehat{V}_j$ and $\widehat{V}_{jk}$ using the estimates from steps 1 and 2 (see \eqref{eq:CredEstVj} and \eqref{eq:CredEstVjk}). Calculate \(\widehat{\widetilde{U}}_{j} = \widehat{V}_j / \hat{\mu}\) and \(\widehat{\widetilde{U}}_{jk} = \widehat{V}_{jk} / \widehat{V}_j\) \; \label{Step3}
	}
	\caption{\label{algo:GLMC}Iterative GLMC algorithm \citep{Ohlsson2008}}
\end{algorithm}

We initialize the model by setting $\widehat{\widetilde{U}}_j = \widehat{\widetilde{U}}_{jk} = 1$ and proceed to the first step, where we fit a GLM. When fitting the GLM, we include $\log(\widehat{\widetilde{U}}_j)$ and $\log(\widehat{\widetilde{U}}_{jk})$ as offset variables and the $w_{ijkt}$'s as weights. This results in the GLM estimates $\hat{\mu}$ (intercept), $\bs{\hat{\beta}}$ (company-specific parameter vector) and $\hat{p}$ (the power parameter). We compute $\hat{\tilde{\mu}} = e^{\hat{\mu}}$ and $\hat{\gamma}_{ijkt} = e^{\bs{x}_{ijkt}^\top \boldsymbol{\hat{\beta}}}$ to proceed to the second step. Here, we first transform the response variable $Y_{ijkt}$ and weight $w_{ijkt}$ as

\begin{equation}
	\begin{aligned}
		\tilde{Y}_{ijkt} &= \frac{Y_{ijkt}}{\gamma_{ijkt}} \ \ \ \text{and} \ \ \ \tilde{w}_{ijkt} &= w_{ijkt} \gamma_{ijkt}^{(2 - p)}.
	\end{aligned}
\end{equation}

\noindent

Consequently, 
\begingroup
\begin{align}
	E[\tilde{Y}_{ijkt}| V_j, V_{jk}] &= \frac{1}{\gamma_{ijkt}} \gamma_{ijkt} V_{jk} = V_{jk}, \nonumber\\ 
	E[\tilde{Y}_{ijkt}| V_j] &= \frac{1}{\gamma_{ijkt}} \gamma_{ijkt} V_{j} = V_{j},\nonumber\\ 
	E\left[\Var{\tilde{Y}_{ijkt}| V_j, V_{jk}}\right] &= E\left[ \frac{1}{\gamma_{ijkt}^2} \frac{\phi \cdot (\gamma_{ijkt} V_{jk})^p}{w_{ijkt}} \right]\\ 
	&= \frac{\phi \cdot E[(V_{jk})^p]}{w_{ijkt} \gamma_{ijkt}^{(2 - p)}} = \frac{\sigma^2}{\tilde{w}_{ijkt}}.\nonumber \nonumber		
\end{align}
\endgroup
\noindent
where $\sigma^2 = \phi \cdot \textcolor{black}{E[(V_{jk})^p]}$. $\tilde{Y}_{ijkt}$ and $\tilde{w}_{ijkt}$ now satisfy the assumptions of the hierarchical credibility model (see Assumption \ref{ass:JewellAssumptions}), thereby enabling us to estimate the variance parameters and to calculate $\widehat{V}_j$ and $\widehat{V}_{jk}$ using equations \eqref{eq:CredEstVj} and \eqref{eq:CredEstVjk}. In the third step, we compute the random effect estimates \(\widehat{\widetilde{U}}_{j} = \widehat{V}_j / \hat{\mu}\) and \(\widehat{\widetilde{U}}_{jk} = \widehat{V}_{jk} / \widehat{V}_j\) using the estimates from steps 1 and 2. Steps 1 to 3 are repeated until the algorithm has converged. \textcolor{black}{Once converged, we predict the damage rate as $\widehat{Y}_{ijkt} = \hat{\tilde{\mu}} \ \hat{\gamma}_{ijkt} \ \widehat{\widetilde{U}}_{j} \ \widehat{\widetilde{U}}_{jk}$.}

\textcolor{black}{Similarly to the hierarchical credibility model, Ohlsson's GLMC algorithm is relatively easy to implement, computationally light and only requires estimates of the mean and variance parameters to obtain the random effect estimates. Further, Ohlsson's approach allows for statistical inference on the parameters estimated by the GLM. Hereto, we use the fitted GLM from the last run in Algorithm \ref{algo:GLMC} and base the inference on the following likelihood}
\begin{equation}\label{eq:logLikOhlsson}
\begin{aligned}
	\prod_j \prod_{k} \prod_{i, t} f (Y_{ijkt} | \hat{\mu}, \hat{\bs{\beta}}, \hat{p}, \log(\widehat{\widetilde{U}}_j), \log(\widehat{\widetilde{U}}_{jk}), \phi, w_{ijkt})
\end{aligned}
\end{equation}
\noindent
\textcolor{black}{where the log-transformed random effects $\widehat{\widetilde{U}}_j$ and $\widehat{\widetilde{U}}_{jk}$ enter as constants. Hence, the statistical inference on $\hat{\mu}$ and $\hat{\bs{\beta}}$ ignores the variability in the random effects.}


\subsubsection{Mixed models}
\textcolor{black}{(Generalized) Linear mixed models (GLMMs) are considered an extension of (G)LMs to the case where responses are correlated or clustered \citep{Molenberghs2005,Tuerlinckx2006}. They are founded in a well-developed statistical framework that provides us with the appropriate inferential tools. The framework of mixed models encompasses a wide range of model specifications, including models with hierarchically structured random effects.}
	
\textcolor{black}{Applied to our setting, the general equation for a mixed model is the same as in equation \eqref{eq:GeneralEquation}.} We assume that $Y_{ijkt} | U_j, U_{jk} \sim \mathcal{E}(g^{-1}(\zeta_{ijkt}), \frac{\phi}{w_{ijkt}} V(g^{-1}(\zeta_{ijkt})))$, where $\mathcal{E}$ denotes any member of the exponential family, and make a distributional assumption on the random effects $U_j$ and $U_{jk}$. In most applications we assume that $U_j \sim \mathcal{N}(0, \sigma^2_{I})$, $U_{jk} \sim \mathcal{N}(0, \sigma^2_{B})$ \citep{REvariance2, NormalREs}, where $\mathcal{N}$ denotes the normal distribution. Verifying these assumptions, however, is often not straightforward. The linear mixed model (LMM) is a special case of a GLMM, where we define $\mathcal{E} \coloneqq \mathcal{N}$ and use the identity-link function $g(\cdot)$. The additive hierarchical credibility model discussed in \eqref{eq:JewellAdditive} is a special case of an LMM and both use the same equations to estimate $\mu$, $ U_j$ and $U_{jk}$ \citep{FreesIME}. The variance parameters, however, are estimated differently in the additive hierarchical credibility model compared to the LMM.

We maximize the marginal likelihood to obtain estimates of the parameters $\mu, \bs{\beta}, \phi, \sigma_{I}^2$, $\sigma_{B}^2$ (and $p$ in case of a Tweedie GLMM). The marginal likelihood is obtained by integrating out the random effects and is given by
\begin{equation}\label{eq:MarginalLikelihood}
	\begin{aligned}
		\prod_j \int \left[ \prod_k \int \prod_{i, t} f(Y_{ijkt} | \Theta, U_j, U_{jk}, \phi, w_{ijkt}) f(U_{jk} | \sigma_{B}^2) d U_{jk} \right] f(U_j | \sigma_{I}^2) d U_j.
	\end{aligned}
\end{equation}
where $\Theta = (\mu, \bs{\beta}, p)$ for a Tweedie GLMM and $\Theta = (\mu, \bs{\beta})$ for other GLMMs. For an LMM, an analytical expression is available for the integrals. In this case, we use the generalized least squares estimator to estimate $\mu$ and $\bs{\beta}$ and rely on either maximum likelihood or restricted maximum likelihood estimators for the estimation of the parameters $\phi$, $\sigma_{I}^2$ and $\sigma_{B}^2$. Conversely, in most GLMMs there is no analytic expression available \textcolor{black}{for the integrals in \eqref{eq:MarginalLikelihood} and we therefore rely on numerical approximations to estimate the parameters.} These approximations can be subdivided into those that approximate the integrand, the data or the integral. A detailed discussion on the different approximation methods is covered in \citet{Molenberghs2005}, \citet{Tuerlinckx2006} and \citet{Frees2014}. \textcolor{black}{In mixed models, we base the statistical inference on \eqref{eq:MarginalLikelihood} and we account for the variability in the random effects when performing inference on $\hat{\mu}$ and $\hat{\bs{\beta}}$. Further, several hypothesis tests are available for the variance parameters $\sigma_{I}^2$ and $\sigma_{B}^2$.}

\textcolor{black}{To predict the realized values of the random effects $U_j$ and $U_{jk}$, we rely on empirical Bayes estimates. Hereto, we base the estimation on the posterior distribution of the random effects given $Y_{ijkt}, \Theta, \phi, w_{ijkt}, \sigma_{I}^2$ and $\sigma_{B}^2$ \citep{Fitzmaurice2014, Molenberghs2005,Skrondal2009}. The density of the posterior distribution of $U_j$ is}
\begin{equation}
	\begin{aligned}
		\propto \prod_k \int \prod_{i, t} f(Y_{ijkt} | \Theta, U_j, U_{jk}, \phi, w_{ijkt}) \ f(U_{jk} | \sigma_{B}^2) \ d U_{jk} \ f(U_j | \sigma_{I}^2).
	\end{aligned}
\end{equation}
\noindent
\textcolor{black}{For $U_{jk}$, the density of the posterior distribution is}
\begin{equation}
	\begin{aligned}
		\propto \int \prod_{i, t} f(Y_{ijkt} | \Theta, U_j, U_{jk}, \phi, w_{ijkt}) \ f(U_j | \sigma_{I}^2) \ d U_j \ f(U_{jk} | \sigma_{B}^2).
	\end{aligned}
\end{equation}
\noindent
\textcolor{black}{The estimates $\widehat{U}_j$ and $\widehat{U}_{jk}$ are those values for $U_j$ and $U_{jk}$ that maximize the corresponding posterior densities. In these densities, the unknown parameters are replaced by their maximum likelihood estimates. For LMMs, we have a closed-form solution for $\widehat{U}_j$ and $\widehat{U}_{jk}$. Conversely, for most GLMMs we do not have an analytical expression available and we have to rely on numerical approximations. Hereafter, we predict the damage rate using}
\begin{equation}
	\begin{aligned}
		\widehat{Y}_{ijkt} = g^{-1}(\hat{\mu} + \bs{x}_{ijkt}^\top \bs{\hat{\beta}} + \widehat{U}_j + \widehat{U}_{jk}).
	\end{aligned}
\end{equation}
\noindent

\subsection{Computational aspects and implementation in R}
We perform our estimations with the statistical software \texttt{R} \citep{Rsoftware}. To estimate the random effects model with the hierarchical credibility model \citep{JewellModel} and the combination of the hierarchical credibility model with a GLM \citep{Ohlsson2008}, we developed our own package called \texttt{actuaRE}. This package is publicly available on \url{https://www.github.com}. For mixed models, a multitude of software implementations are available alongside with detailed documentation. In our paper, we rely on the \citePkg{lme4} and \citePkg{cplm} packages to estimate the random effects model using the mixed model framework.

Estimation via the hierarchical credibility model, as discussed in Section \ref{Sec:Jewell}, is fastest in terms of computation time. Estimation via (G)LMMs is by far the slowest as they require the approximation and maximization of complicated likelihoods. Computationally, GLMMs are complex and they are more likely to experience convergence problems (see \citet{ConvergenceWarnings} for information on how to handle convergence warnings). Related hereto is that, in certain situations, we may obtain negative variance estimates and this may occur for all estimation methods. Within the mixed models framework, this is a well-known problem \citep{Pryseley2011}. Negative variance estimations may be due to low variability \citep{Oliveira2017} or a misspecification of the hierarchical MLF \citep{Pryseley2011}.



\section{Case study: workers' compensation insurance}\label{sec:3}
We illustrate the predictive model building work flow on a workers' compensation insurance data set from a Belgian insurer. In this data set, we have a hierarchical MLF and company-specific covariates at our disposal. Further, we have the company identification number for each of the companies in the portfolio which enables us to retrieve company-specific financial information from an external data source. We refer to the externally acquired data as the external database. To preserve the confidentiality of the data, we omit all confidential information. Hereto, we either remove all values from the figures or apply a transformation when showing values.


\subsection{Internal data set}\label{sec:InternalDataset}
To prevent that large claims distort our findings, we start the analysis by capping large claim amounts $Z_{hijkt}$ using concepts from extreme value theory (EVT) \citep{StatExtremes}. Here, the index $h$ refers to an individual claim of company $i$ operating in branch $k$ within industry $j$ in year $t$. In the analysis we use the ratio of $Z_{hijkt}$ to a year-specific correction factor $c_{t}$, thereby accounting for inflation. Using tools from EVT, we determine the threshold $\tau$ between the attritional losses and the large losses. We transform $\tau$ to a year-specific threshold using $\tau_t = \tau \times c_t$ and cap $Z_{ijkt}$ as follows
\begin{equation}
	\begin{aligned}
		\tilde{Z}_{hijkt} = \text{min}(Z_{hijkt}, \tau_{t})
	\end{aligned}
\end{equation}
\noindent
where $\tilde{Z}_{hijkt}$ denotes the capped claim amount. Thereafter, we redistribute the total capped amount among all claims based on their share in the total cost. Hereby, we ensure that the total claim amount after redistribution equals the total claim amount before capping. Given the confidentiality of the data, we do not disclose how we redistribute the total capped amount.

\begin{figure}[!htbp]
	\centering
	\caption{\label{fig:EmpDistr}Empirical distribution of the damage rates $Y_{ijkt}$ of the individual companies}
	\makebox[\textwidth][c]{\includegraphics[scale = 0.5]{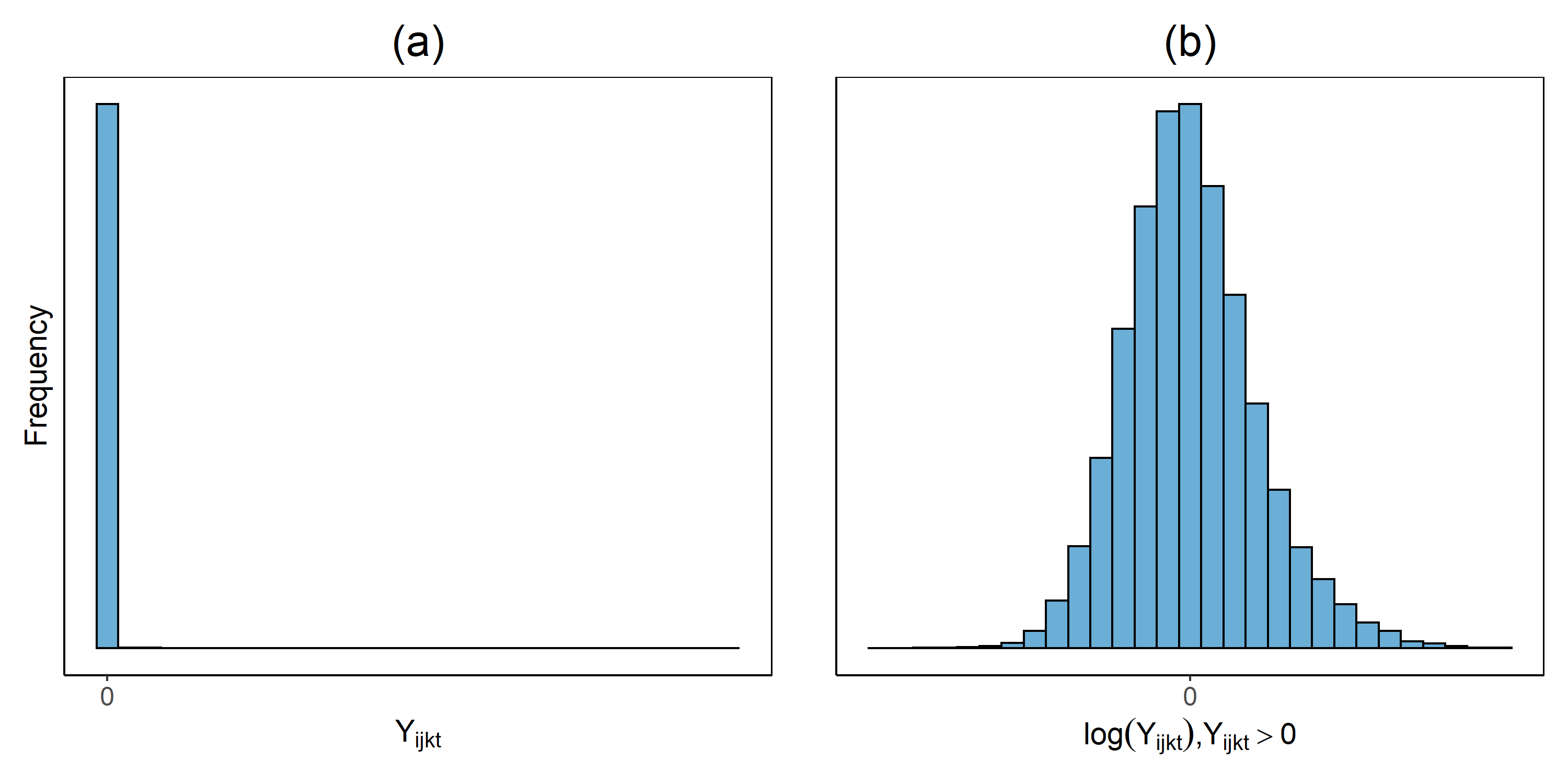}}
\end{figure}

After this first data preprocessing step, we compute the damage rate for each of the individual companies as
\begin{equation}\label{eq:DamageRate}
	\begin{aligned}
		Y_{ijkt} = \frac{\sum_h \mathcal{Z}_{hijkt}}{w_{ijkt}}.
	\end{aligned}
\end{equation}
\noindent
The empirical distribution of the damage rates $Y_{ijkt}$ of the individual companies is visualized in panel (a) of \Cref{fig:EmpDistr}. The empirical distribution is characterized by a strong right skew and this right skew is still present when log transforming $Y_{ijkt}$ for $Y_{ijkt} > 0$ (see \Cref{fig:EmpDistr}\textcolor{red}{(b)}). Of all the individual damage rates $Y_{ijkt}$, 85\% equals zero and a mere 0.4\% of the $Y_{ijkt}$'s are larger than one (\Cref{fig:EmpDistr}\textcolor{red}{(a)}). Hence, the majority of the damage rates are either zero or relatively low compared to the salary mass.

\begin{figure}[!ht]
	\centering
	\caption{\label{fig:TreeMap}Tree maps depicting hierarchical structure}
	\makebox[\textwidth][c]{\includegraphics[scale = 0.4]{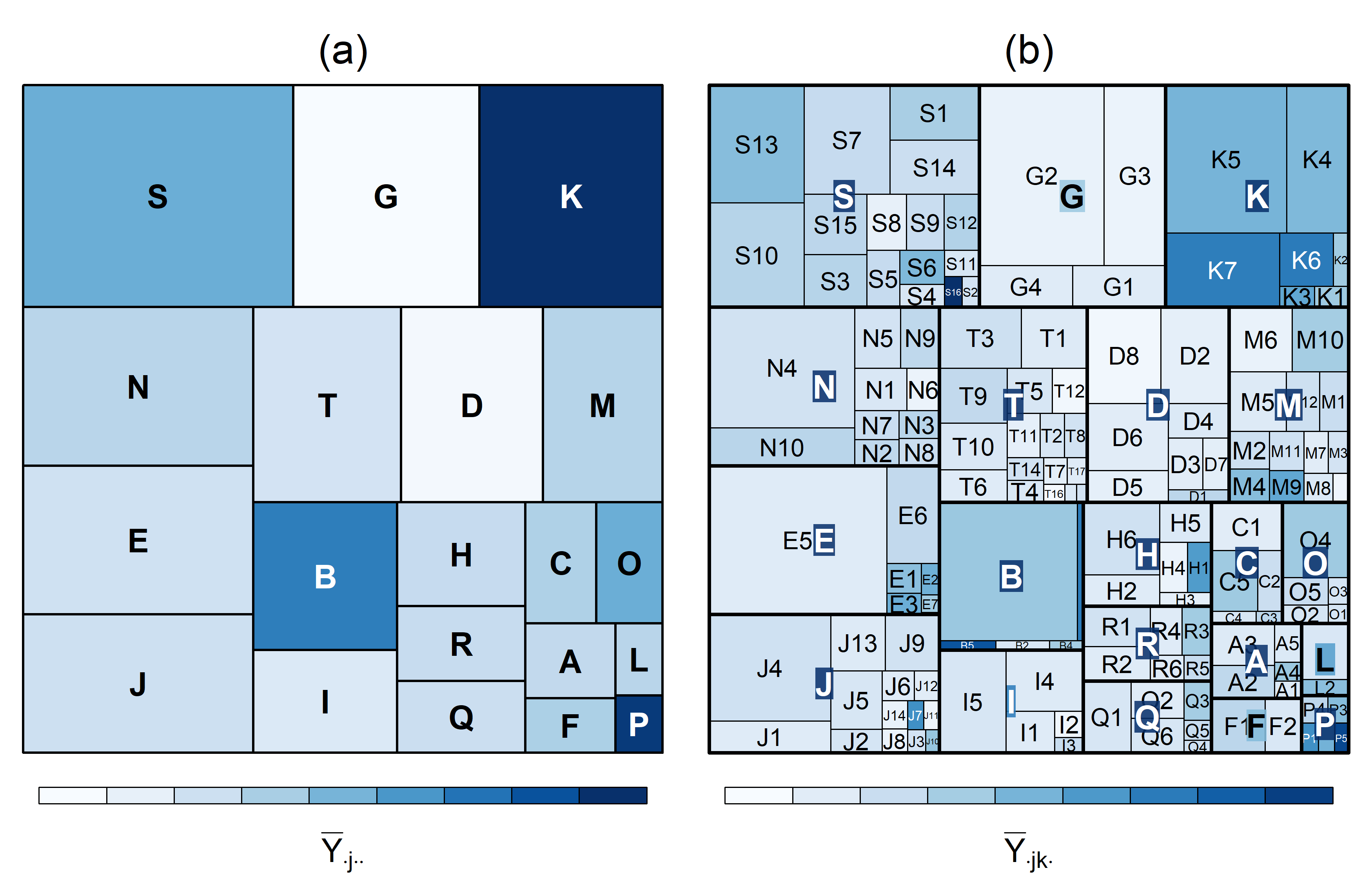}}
\end{figure}

We use the individual $Y_{ijkt}$ and corresponding $w_{ijkt}$ to compute the weighted average of the damage rates at the industry- and branch-level 
\begin{equation}
	\begin{aligned}
		\bar{Y}_{\cdot j \cdot \cdot} = \frac{\sum_{i, k, t} w_{ijkt} Y_{ijkt}}{\sum_{i, k, t} w_{ijkt}}, \mspace{75mu} 
		\bar{Y}_{\cdot j k \cdot} = \frac{\sum_{i, t} w_{ijkt} Y_{ijkt}}{\sum_{i, t} w_{ijkt}}
	\end{aligned}
\end{equation}
\noindent
and visualize these in the treemaps in \Cref{fig:TreeMap}. Panel (a) shows the $\bar{Y}_{\cdot j \cdot \cdot}$'s and panel (b) the $\bar{Y}_{\cdot j k \cdot}$'s. In the treemaps, the summed salary mass of the industries and branches within industries determines the size of the rectangles and a color gradient is used for the weighted averages. The larger the summed salary mass, the larger the rectangle and the larger the weighted average, the darker the color.

Considerable variation is present between industries as well as within industries in the weighted averages. In addition, we see that the $\bar{Y}_{\cdot j k \cdot}$'s are more similar within industries than between industries (see Figure \ref{fig:TreeMap}\textcolor{red}{(b)}). For industry K, for example, the $\bar{Y}_{\cdot j k \cdot}$'s are visibly larger than those of industry D and within industry K, there are clear differences between the different branches (e.g. between K1 and K7). Consequently, an indispensable part of the variation of $Y_{ijkt}$ seems to be attributable to the industry and branch in which the companies operate. 

In addition to the hierarchical MLF, we have company-specific covariates at our disposal, such as the number of FTEs or the company type. We refer to these covariates as the internal variables. For a particular level $l$ of a covariate, we calculate the weighted average of the damage rates as
\begin{equation}\label{eq:EmpWeightedAvg}
	\begin{aligned}
		\bar{Y}_{l} = \frac{\sum\limits_{(i, j, k, t) \mspace{5mu} \in \mspace{5mu} l} w_{ijkt} Y_{ijkt}}{\sum\limits_{(i, j, k, t) \mspace{5mu} \in \mspace{5mu} l} \ w_{ijkt}}.
	\end{aligned}
\end{equation}
\noindent
Here, $\sum_{(i, j, k, t) \mspace{5mu} \in \mspace{5mu} l}$ indicates that the summation is limited to those observations that are categorized into level $l$. For example, when computing $\bar{Y}_{l}$ with $l = $ \texttt{1} for \texttt{internal variable 1}, we only consider the $Y_{ijkt}$'s of companies that have the value \texttt{1} for \texttt{internal variable 1}. By comparing the $\bar{Y}_{l}$ across the different levels we empirically explore whether certain levels are considered to be more risky than others with a marginal, empirical analysis.

\begin{figure}[!ht]
	\centering
	\caption{\label{fig:DescrInternal}Comparison of the $\bar{Y}_{l}$'s per covariate}
	\includegraphics[width = \textwidth]{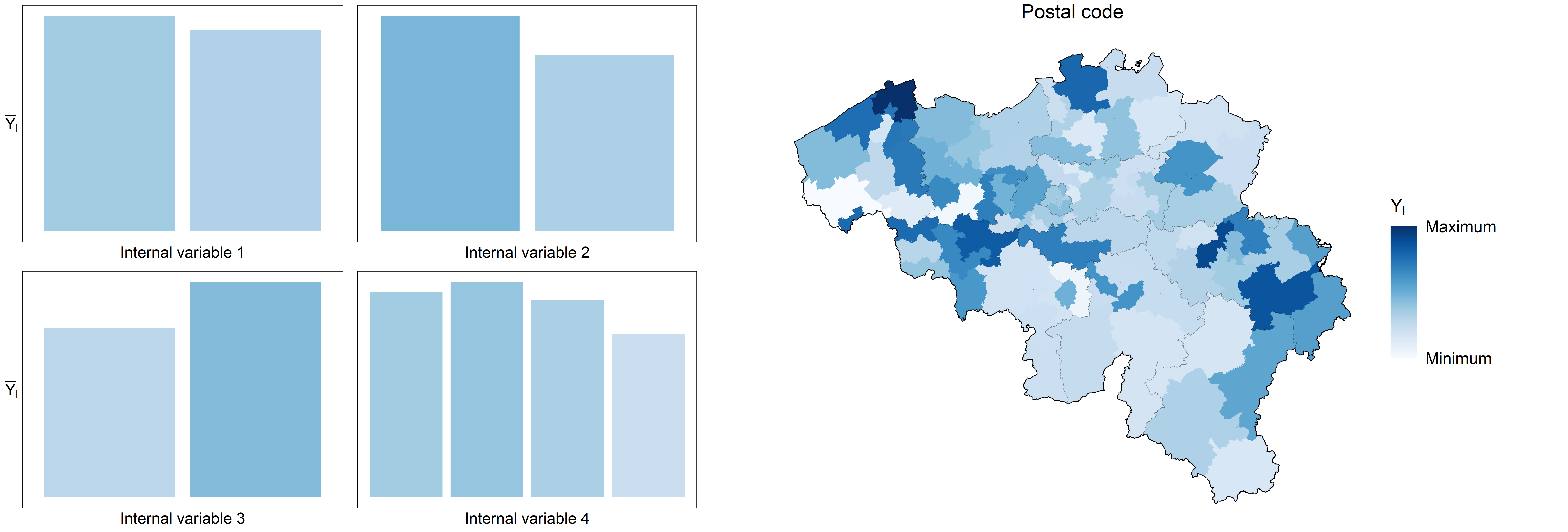}
\end{figure}

\Cref{fig:DescrInternal} shows the $\bar{Y}_{l}$'s for the internal variables. To preserve the confidentiality of our findings, we randomly allocate the postal codes to different regions on the map. We preserve this allocation throughout the article for consistency of the results. The weighted average mainly differs between the levels of the variables \texttt{internal variable 2}, \texttt{internal variable 3}, \texttt{internal variable 4} and \texttt{postal code}.



\subsection{External data set}
The second source of information is the Bel-First database (\url{https://belfirst.bvdinfo.com}) which contains the financial statements of all Belgian private entities that file their financial accounts to the National Bank of Belgium \citep{BelFirst}. For each of the companies, we have yearly data available in the Bel-First database. We link the claims observed during year $t$ with the financial performance indicators for year $t-1$. For example, we extract the financial information of company $i$ in year 2019 and link this to the registered claims of company $i$ in 2020.

We retrieve information on 30 variables, 26 of which are related to the financial situation of the company and we are able to retrieve information for approximately 70\% of the companies. For all extracted variables, we have occasional missing values. For 30\% of the companies we are not able to retrieve any financial information since these companies are not obliged to report to the National Bank of Belgium. Of this 30\%, the majority of the companies are categorized as independent natural persons, craftsmen. A minority of this 30\% are non-profit organizations, private companies with a limited liability, Belgian companies/associations without accounts or Belgian companies/associations that do not file their account in a standard model.


Using the available financial information, additional variables are created. We specify a binary variable that indicates whether the company is considered to be a zombie firm or not and we use the definition of \citet{Adalet2018}. According to \citet{Adalet2018}, a firm is identified as a zombie firm when its interest coverage ratio (ICR) has been less than one for at least three consecutive years and if the company is at least 10 years old. Next to this zombie variable, we compute the relative change of the variables that are commonly associated with growth. These variables are sales, number of employees, total assets, cash flow and added value (see, for example, \citet{Vanacker2008}). The relative change at time $t$ for a variable $X$ is then computed as $(X_{t} - X_{t - 1}) /|X_{t - 1}|$.

\subsection{Binning continuous and spatial company-specific covariates}\label{sec:Binning}
In order to arrive at an interpretable tariff list that is easily explainable to all stakeholders, we transform continuous and spatial company-specific covariates to categorical ones. Hereto we employ a data-driven binning strategy based on the work of \citet{Henckaerts2018}. We use a different strategy for continuous and spatial variables to account for the variable type. For continuous variables, we want to preserve the ordering and only allow for the binning of consecutive values. Conversely, for spatial variables we want a strategy that enables us to merge non-adjacent postal codes. In this section, we provide a general description of the binning process.

\paragraph*{Continuous variables.} We start the binning process of a continuous variable by fitting a univariate generalized additive model (GAM) to the company-level data. In this exploratory preprocessing step, we do not include any random effects in the GAM for computational simplicity and fit the following model
\begin{equation}
	\begin{aligned}
		g(E[Y_{ijkt} | {}_{BF} x_{ijkt}, x_{ijkt}]) = \mu + {}_{BF} x_{ijkt} \beta_{BF} + I(x_{ijkt} \ \text{available}) f(x_{ijkt})
	\end{aligned}
\end{equation}
\noindent
where $i$ serves as an index for the company. ${}_{BF} x_{ijkt}$ is a binary variable indicating that no financial information is available for company $i$ in the Bel-First database, $x_{ijkt}$ is the external variable, $I(x_{ijkt} \ \text{available})$ indicates whether $x_{ijkt}$ is known ($I(x_{ijkt} \ \text{available}) = 1$) or not ($I(x_{ijkt} \ \text{available}) = 0$) and $f(\cdot)$ denotes the smooth effect. This model specification allows us to use all available external information and we hereby do not omit information from companies that either cannot be found in the Bel-First database or that can be found in the Bel-First database, but have a missing value for the covariate. Missing values are assumed to be missing at random for companies found in the Bel-First database. Given the size of our data set we opt for the simplicity of the indicator method to handle missing data \citep{Bennett2001}. In addition, we include ${}_{BF} x_{ijkt}$ as a confounding variable to account for the fact that certain companies are not found in the Bel-First database. The wage bills $w_{ijkt}$ are incorporated as weights. To examine the effect of the distributional assumption, we perform this procedure once using univariate GAMs with a Gaussian distribution and identity link and once using univariate GAMs with a Tweedie distribution and log link. 

\begin{figure}[!ht]
	\centering
	\caption{\label{fig:ExampleBinningContinuous}Illustration of the binning process for continuous covariates. The histogram on the left shows the empirical distribution of the variable \texttt{net added value}, after limiting its range to non-outlying values. The figure on the right depicts the fitted smooth effect $g^{-1}(\mu + \hat{f}(x_{ijkt}))$ (black solid line) together with the 95\% confidence interval (black dashed lines). Here, the blue bars depict the empirical weighted averages by consecutively grouping values until they contain at least 5\% of the observations.}
	\makebox[\textwidth][c]{\includegraphics[width = \textwidth]{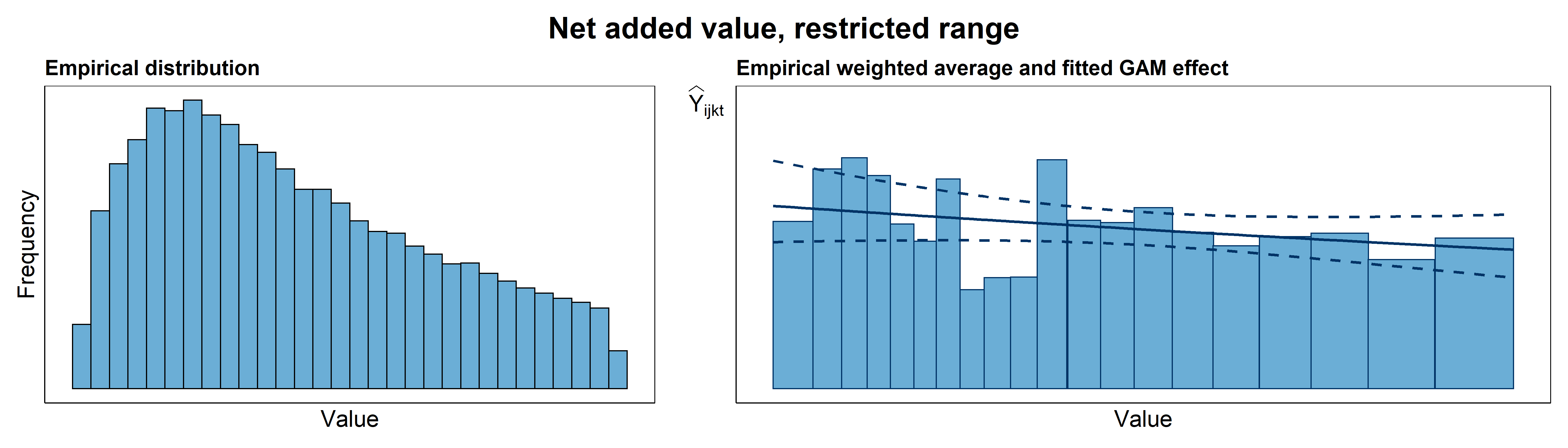}}
\end{figure}

An illustration of the binning process for the variable \texttt{net added value} is given in \Cref{fig:ExampleBinningContinuous} when assuming a Gaussian distribution for the response in the GAM. The histogram on the left side of the figure shows the empirical distribution of the continuous variable (after limiting its range to non-outlying values to focus on the pattern seen in the majority of the companies) and the figure on the right side shows the fitted smooth effect $g^{-1}(\mu + \hat{f}(x_{ijkt}))$ (black solid line) together with the 95\% confidence interval (black dashed lines). The blue bars on the right side depict the empirical weighted averages by consecutively grouping values until they contain at least 5\% of the observations.

The fitted smooth effect is binned with a regression tree and the resulting bins are inspected in detail for every covariate. In addition to inspecting the fitted smooth effect and the resulting bins at the original scale of the covariate, we assess the binning based on the log-transformed counterpart for positively valued covariates and we choose the binning that best approximates the empirical weighted averages. 

\begin{figure}
	\makebox[\linewidth][c]{%
		\caption{Results binning two-digit postal code}
		\begin{subfigure}[b]{.5\textwidth}
			\centering
			\caption{\label{fig:GausBinPC}Gaussian assumption}
			\includegraphics[width=.95\textwidth]{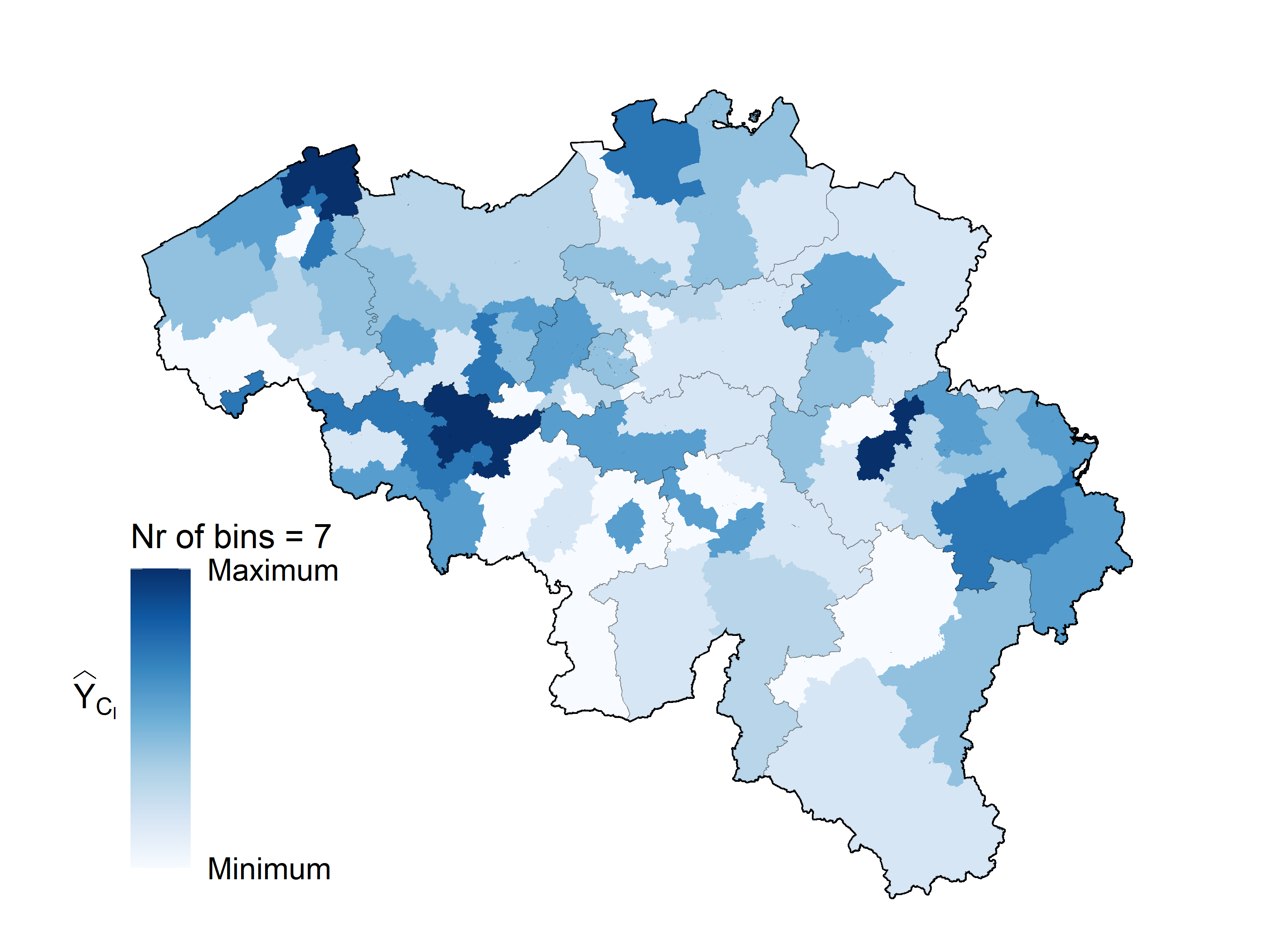}
		\end{subfigure}%
		\begin{subfigure}[b]{.5\textwidth}
			\centering
			\caption{\label{fig:TweedieBinPC}Tweedie assumption}
			\includegraphics[width=.95\textwidth]{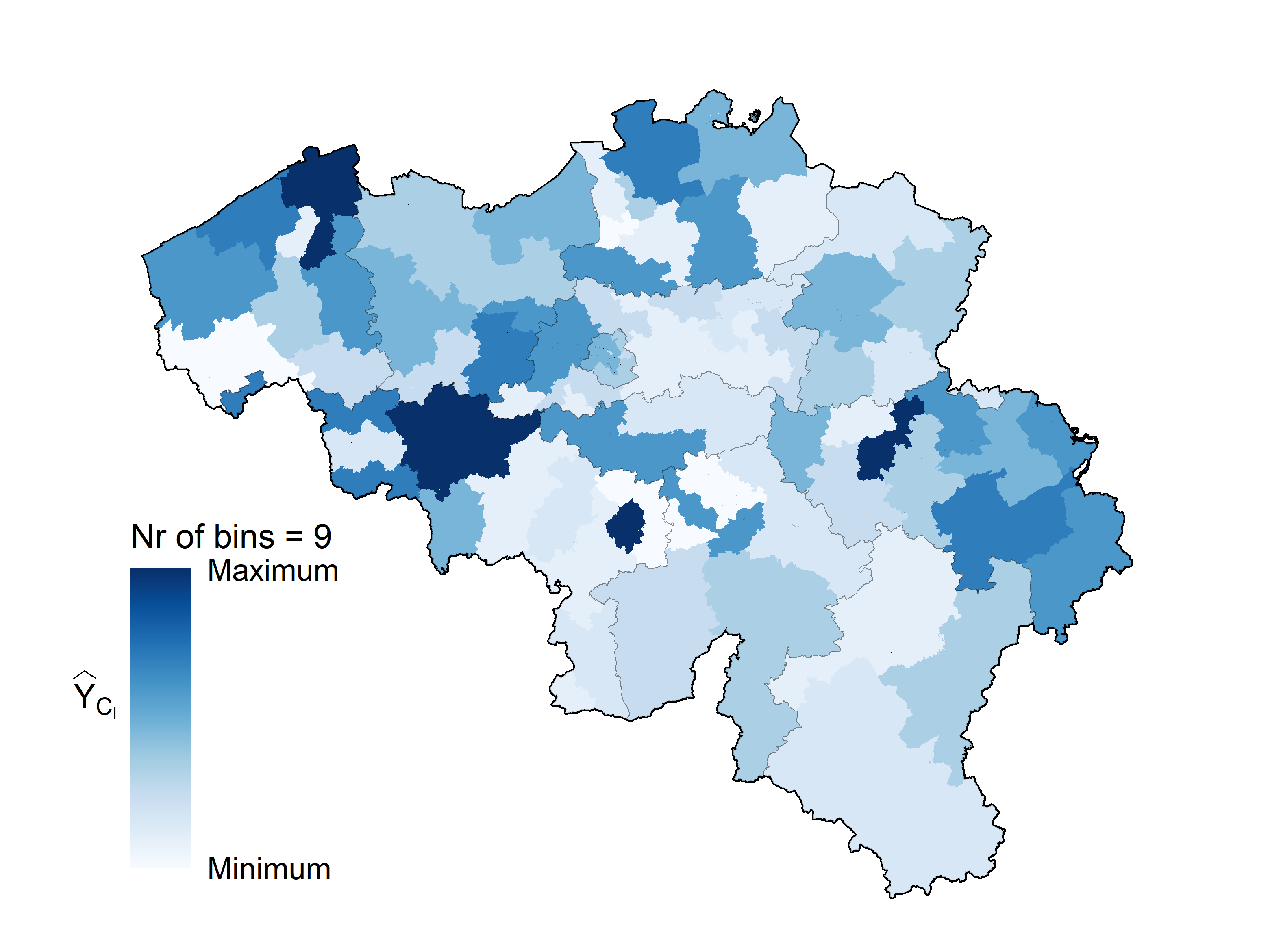}
		\end{subfigure}%
	}\\
\end{figure}

\paragraph*{Spatial variable.} For postal code, we first construct preliminary clusters by using only the first two digits of the postal code. We use dummy variables to encode the two-digit postal code and fit the following model to the company-level data
\begin{equation}
	\begin{aligned}
		g(E[Y_{ijkt} | U_j, U_{jk}]) &= \mu  + \boldsymbol{x}_{ijkt}^\top \boldsymbol{\beta} + U_j + U_{jk}.\\
	\end{aligned}
\end{equation}
\noindent
where the covariate vector $\boldsymbol{x}_{ijkt}$ consists of the dummy variables. The model is fit using Ohlsson's iterative algorithm (see Algorithm \ref{algo:GLMC}) and the estimated coefficients are clustered using the Ckmeans.1d.dp algorithm \citep{BinningPC}. To select the number of clusters $n_c$, we perform a grid search with the AIC as criterion. The results of the clustering strategy are shown in \Cref{fig:GausBinPC,fig:TweedieBinPC} when assuming a Gaussian and Tweedie distribution for the response, respectively. The Gaussian model specification results in seven separate clusters and the Tweedie model specification results in nine clusters. After clustering, we refit the model using Ohlsson's algorithm. The colors in the plot depict the estimated damage rate for each of the clusters. We calculate the estimated damage rate for cluster $C_l$ as $\widehat{Y}_{C_l} = g^{-1}(\mu  + \widehat{\beta}_{C_l})$ for $l = (1, \dots, n_c)$. Here, $\widehat{\beta}_{C_l}$ denotes the estimated coefficient for cluster $C_l$. Overall, both results closely resemble each other and are in line with the results of our exploratory analysis shown in Figure \ref{fig:DescrInternal}. 


\subsection{Development of the predictive model}\label{sec:DevelopmentPredModel}
We split the data into a training and test set. We use the training set to develop the predictive model and the test set to assess its predictive performance. The training set contains data from the first seven years and the test set contains data from the eighth and most recent year.



\paragraph*{Preselection of the external covariates.} Considering that we have a substantial amount of externally selected company-specific covariates, we first perform a preliminary variable selection to retain the most important external covariates. Since this step determines which external covariates are investigated further, we rely on the well-developed statistical framework of mixed models. This framework enables us to accurately estimate the external covariate parameters and to calculate the marginal AIC (mAIC) of the fitted GLMM \citep{Saefken2014}. The mAIC focuses on the fixed effects and we use this criterion to select the covariates that fit the data well. We start by fitting univariate GLMMs (i.e. only one external covariate is included) to the company-level data with the following general equation
\begin{equation}\label{eq:UnivariateGLMM}
	\begin{aligned}
		g(E[Y_{ijkt} | U_j, U_{jk}]) = \mu + {}_{BF} x_{ijkt} \beta_{BF} + {}_{Ext}\bs{x}_{ijkt}^\top  \bs{\beta}_{Ext}  + U_j + U_{jk}.
	\end{aligned}
\end{equation}
\noindent
Here, ${}_{Ext}\bs{x}_{ijkt}$ consists of the dummy variables for the binned external covariate as obtained from \hyperref[sec:Binning]{Section 3.3} and $\bs{\beta}_{Ext}$ denotes the corresponding parameter vector. To make the model identifiable, observations with a missing value for the external covariate serve as a reference. We compute the mAIC of the univariate GLMM as specified in \eqref{eq:UnivariateGLMM} using
\begin{equation}
	\begin{aligned}
		\text{mAIC} = -2 \log(f(Y_{ijkt} | \Theta, \phi, \sigma_{I}^2, \sigma_{B}^2)) + 2 (n_p + q + 1)
	\end{aligned}
\end{equation}
\noindent
where $\log(f(Y_{ijkt} | \Theta, \sigma^2, \sigma_{I}^2, \sigma_{B}^2)$ denotes the marginal log-likelihood (see equation \eqref{eq:MarginalLikelihood}), $\Theta = (\mu, \beta_{BF}, \bs{\beta}_{Ext}, p)$ for Tweedie GLMMs and $\Theta = (\mu, \beta_{BF}, \bs{\beta}_{Ext})$ for other GLMMs, $n_p$ the number of parameters of the external covariate plus intercept and $q$ the number of variance parameters of the random effects. Note that we add a one to $n_p + q$ to account for the estimation of the dispersion parameter $\phi$. In case of a Tweedie GLMM, we add a two to $n_p + q$ to account for the estimation of the dispersion parameter $\phi$ and the power parameter $p$. We use the mAIC values to select the top 5 covariates. \textcolor{black}{We rely on the mAIC, since information criteria are better suited for model selection than statistical tests \citep{Burnham2007}.}

This procedure results in a different set of preselected external covariates, depending on the assumed distribution for the response. Comparing the results based on the LMM and Tweedie GLMM, we see that \texttt{external variable 2} and \texttt{external variable 3} are the only variables that are selected by both models. 

Using the internal and preselected external variables, we compute the damage rate $Y_{ijkt}$ for each possible combination of company-specific covariates and hierarchical MLF values and one such combination determines a tariff class. We use $i$ as an index for the tariff class and calculate $Y_{ijkt}$ as
\begin{equation}
	\begin{aligned}
		Y_{ijkt} = \frac{\sum_h \mathcal{Z}_{hijkt}}{\sum_h w_{hijkt}}
	\end{aligned}
\end{equation}
\noindent
where $\mathcal{Z}_{hijkt}$ refers to the capped claim amount of the $h^{th}$ company in tariff class $i$ operating in branch $k$ within industry $j$ at time $t$ and $\sum_h w_{hijkt}$ represents the sum of the corresponding salary masses. All possible tariff classes are combined into a tariff table.

\paragraph*{Variable selection.} Next, we apply best subset regression \citep{Subset1, Subset2} with the Akaike Information Criterion (AIC) \citep{AIC} as selection criterion. The general model equation is given by
\begin{equation}\label{eq:EquationVarSel}
	\begin{aligned}
		g(E[Y_{ijkt} | U_j, U_{jk}]) &= \mu + {}_{Int} \bs{x}_{ijkt}^\top \boldsymbol{\beta}_{Int} +  {}_{BF} x_{ijkt} \beta_{BF} + {}_Z x_{ijkt} \beta_Z \\ 
		&+  \leftidx{_{Ext}}{\bs{x}}{_{ijkt}^\top}  \bs{\beta}_{Ext}  + U_j + U_{jk}.\\
	\end{aligned}
\end{equation}
\noindent
where subscripts $Int$ and $Z$ refer to the internal variables and zombie variable, respectively. $\leftidx{_{Ext}}{\bs{x}}{_{ijkt}}$ refers to the covariate vector of the external covariates with corresponding parameter vector $\boldsymbol{\beta}_{Ext}$. The observations with missing values serve as a reference for the zombie and external variables.

To estimate the parameters in equation \eqref{eq:EquationVarSel} we use Ohlsson's GLMC algorithm. We fit models with all possible combinations of the company-specific covariates and include the hierarchical MLF in all models. We opt for Ohlsson's GLMC algorithm given its simplicity and computational efficiency. In comparison, GLMMs are computationally heavy, frequently experience convergence issues and are therefore not suited for exhaustive variable selection methods using this data set. Ohlsson's estimation method, however, does not allow us to calculate the mAIC since this method does not maximize the marginal likelihood. We therefore extract the AIC from the GLM fit resulting from the last iteration in Ohlsson's GLMC algorithm and we select the model with the lowest AIC. \textcolor{black}{Hereby, we select the best fitting parsimonious model from a set of fitted models.}

We first perform best subset regression with the set of internal covariates only and in every model, we include the hierarchical MLF. Following this, we perform best subset regression with the set of external covariates and include the selected internal covariates, ${}_{BF} x_{ijkt}$ and hierarchical MLF in all models. We perform this procedure once with a Gaussian model specification and identity link and once with a Tweedie model specification with a log link.


\Cref{tab:ResultsSubset} shows the results of the variable selection procedure. In the internal covariates only models, the first two internal variables and \texttt{two-digit postal code binned} are selected. When external covariates are entered, \texttt{external variable 2} is selected in the Gaussian model. Conversely, with the Tweedie model specification the first and third external variable are selected.

\begin{table}[!htbp]
	\caption{\label{tab:ResultsSubset}Results best subset regression}
	\footnotesize
	\begin{center}
		\scalebox{0.875}{
			\centering
			\makebox[\textwidth][c]{%
				\begin{tabular}{@{\extracolsep{4pt}}p{1pt}>{\fontfamily{lmr}\selectfont}l*{4}{c}@{}}
					\toprule
					& & \multicolumn{2}{c}{Gaussian} & \multicolumn{2}{c}{Tweedie}\\
					\cline{3-4} \cline{5-6}
					\rule{0pt}{4ex}&\multicolumn{1}{l}{Predictor}   & Internal only  & Internal + external & Internal only  & Internal + external \\ \midrule
					\multirow{5}{*}{\rotatebox{90}{Internal}}
					& \texttt{Internal variable 1}       & $\times$ &  $\times$  &  $\times$  &  $\times$  \\
					& \texttt{Internal variable 2} &  $\times$  &  $\times$  &  $\times$  &  $\times$           \\
					& \texttt{Internal variable 3}             &&&&          \\
					& \texttt{Internal variable 4} &&&&          \\
					& \texttt{Two-digit postal code binned}        &  $\times$  &  $\times$  &  $\times$  &  $\times$             \\
					\midrule
					\multirow{4}{*}{\rotatebox{90}{External}}
					& \texttt{Available information Bel-First} & & $\times$ & & $\times$ \\
					& \texttt{External variable 1} &&&& $\times$ \\
					& \texttt{External variable 2}   && $\times$ &&\\
					& \texttt{External variable 3} &&&& $\times$ \\
					\bottomrule
				\end{tabular}
			}
		}
	\end{center}
\end{table} 

\paragraph*{Benchmark models.} The models resulting from best subset regression allow us to examine the predictive performance when we use internally and externally collected company-specific covariates as well as a hierarchical MLF. We also fit a hierarchical credibility model and an intercept-only (G)LMM to the training set which serve as benchmark models.

\subsection{Inspecting the model fits on the training set}\label{sec:ModelFit}
\textcolor{black}{Following, we refit and refine the selected models. To evaluate the distributional assumption on the target variable and the goodness of fit, we examine the estimated effect sizes of the company-specific covariates and random effects (Section \ref{sec:ModelFit}) as well as the fitted values on the training set (Section \ref{sec:FittedValues}) in detail. An appropriate distributional assumption will provide a good fit with the data. In this case, the estimated values of the company-specific parameters and random effects are expected to be in line with the findings of the exploratory analysis in Section \ref{sec:InternalDataset}.} 

\paragraph*{Examining the company-specific covariates} We refit the models selected by best subset regression, as sketched in Section \ref{sec:3}, using GLMMs and use the 95\% confidence intervals (CIs) of the estimated coefficients of the company-specific covariates to refine the model. For variables with a large number of levels, we need an alternative strategy. Here, we want to reduce the number of levels. Hereto, we use the multi-type Lasso \citep{smurf} with the Fused Lasso penalty for these variables to merge consecutive levels in a data-driven way. To account for the hierarchical structure, we use the random effect estimates of the GLMMs and specify these as offset variables in the multi-type Lasso. We include the salary mass as weight.

\begin{figure}
	\makebox[\linewidth][c]{%
		\caption{Internal covariates only models}
		\begin{subfigure}[b]{.5\textwidth}
			\centering
			\caption{\label{fig:BaseModelLMM}LMM}
			\includegraphics[width=.95\textwidth]{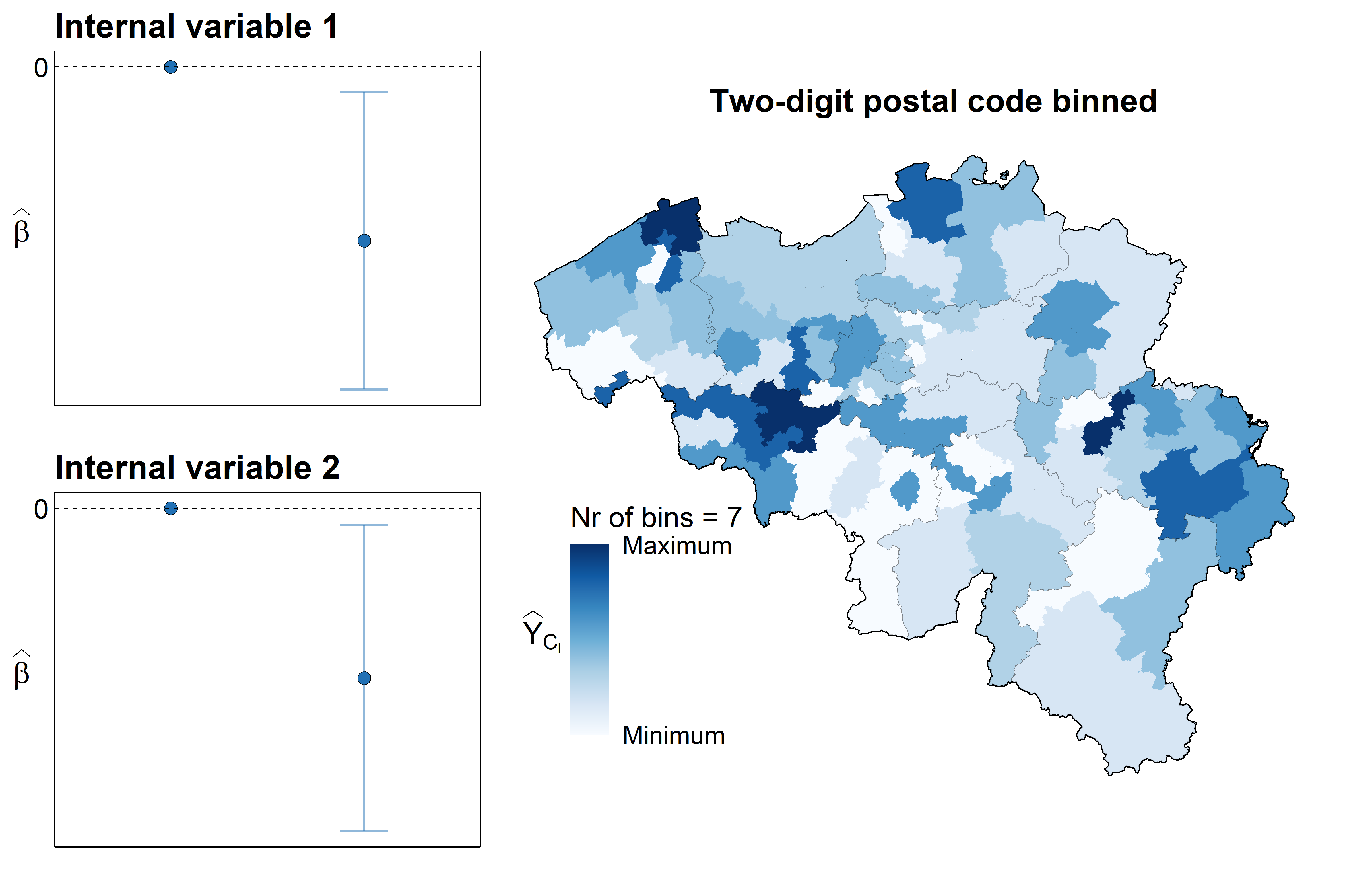}
		\end{subfigure}%
		\begin{subfigure}[b]{.5\textwidth}
			\centering
			\caption{\label{fig:BaseModelTweedie}Tweedie GLMM}
			\includegraphics[width=.95\textwidth]{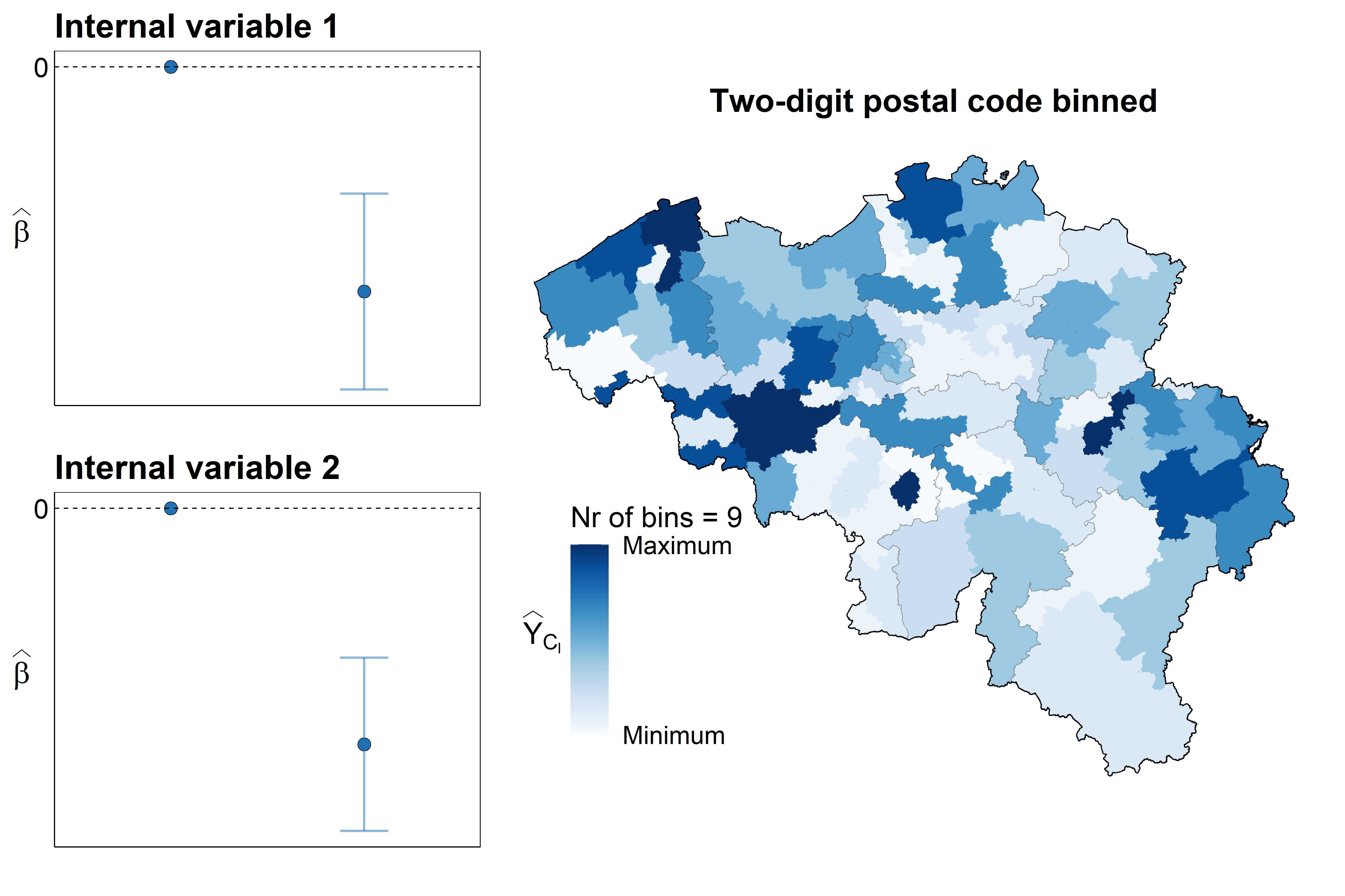}
		\end{subfigure}%
	}\\
\end{figure}

\Cref{fig:BaseModelLMM,fig:BaseModelTweedie} show the estimated coefficients of the company-specific risk factors and the 95\% CIs for the internal covariates only models (i.e. the models resulting from best subset regression with only internal covariates). The direction of the estimated coefficients (positive or negative) is the same for the LMM and Tweedie GLMM. In addition, the model fits confirm the findings of our exploratory analysis.\ignore{Compared to small-sized companies, medium- to large-sized companies are considered to be less risky. Small-sized companies that receive a loyalty code are also considered to be less risky. Conversely, companies that are located in the Southern part of Belgium are considered to be more risky in comparison to companies that are located in the Northern part of Belgium.} Considering that none of the confidence intervals are close to zero, no adjustments are made to the internal covariates only models.

\begin{figure}[!htbp]
	\makebox[\linewidth][c]{%
		\caption{Internal + external covariates models: coefficient estimates external covariates}
		\begin{subfigure}[b]{.5\textwidth}
			\centering
			\caption{\label{fig:FinalAllModelLMM}LMM}
			\includegraphics[scale = 0.25]{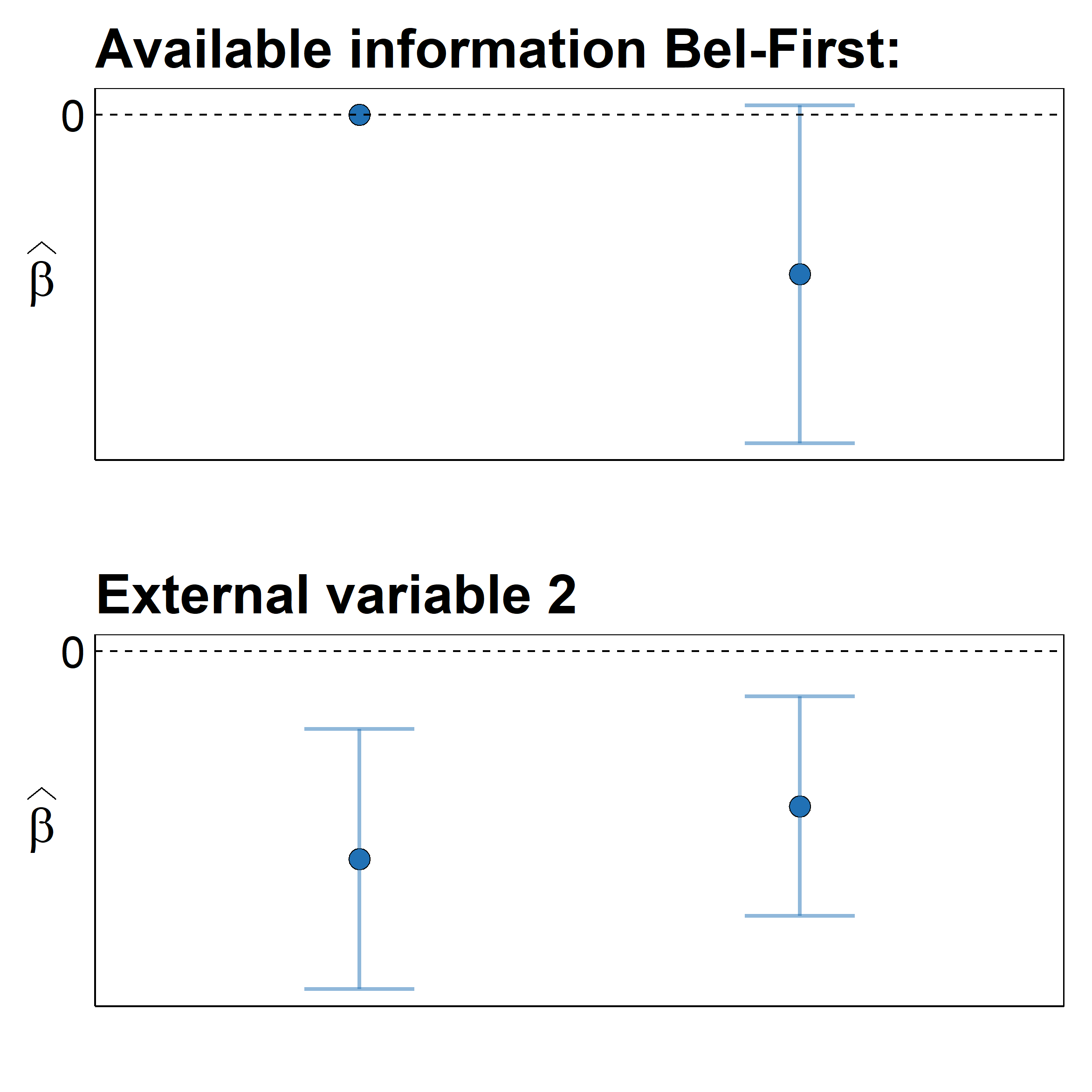}
		\end{subfigure}%
		\begin{subfigure}[b]{.5\textwidth}
			\centering
			\caption{\label{fig:FinalAllModelTweedie}Tweedie GLMM}
			\includegraphics[scale = 0.25]{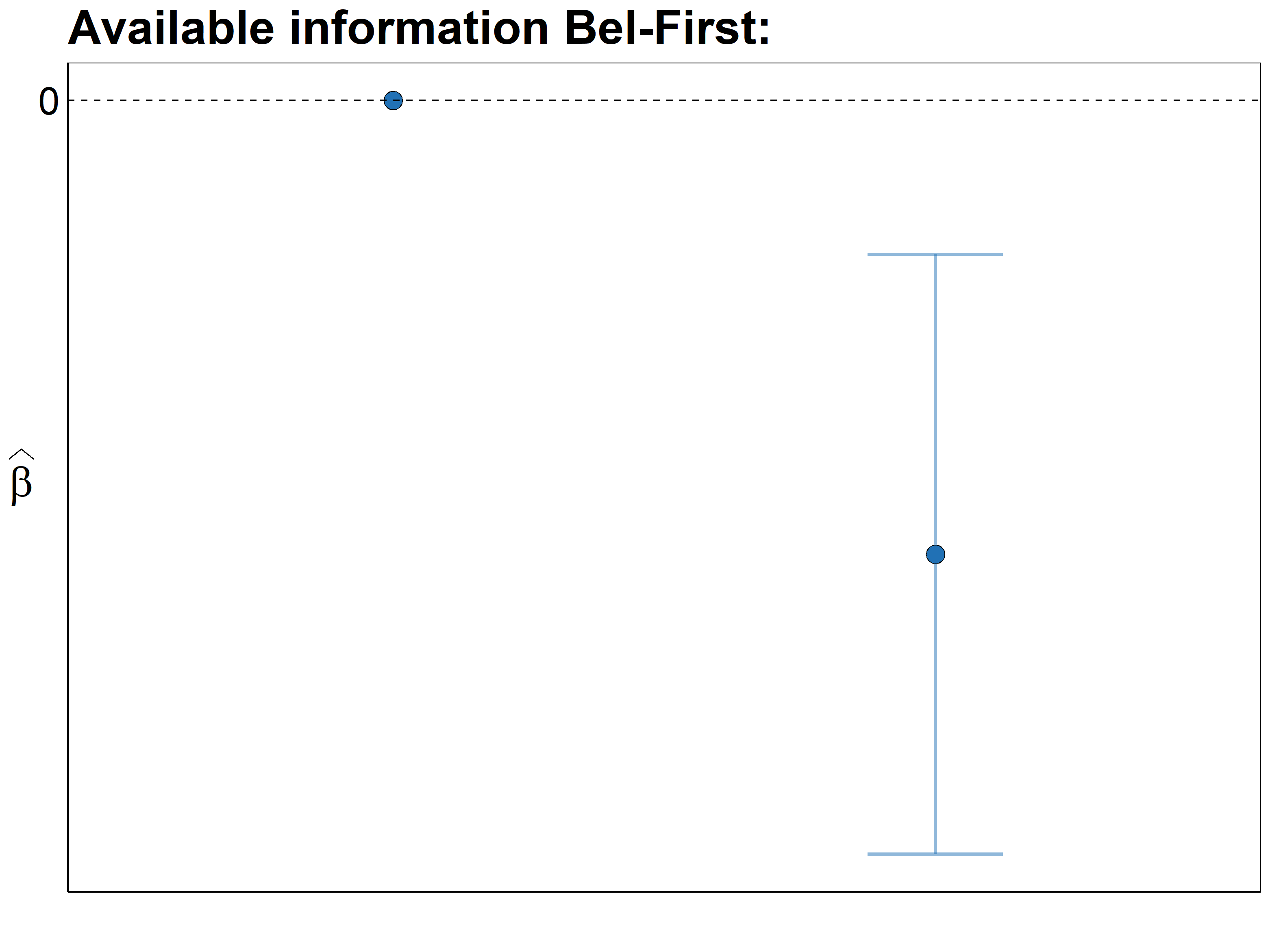}
		\end{subfigure}%
	}\\
\end{figure}

When adding external covariates to the internal covariates only LMM, \texttt{external variable 2} is selected by best subset regression. For the Tweedie GLMM, the covariates \texttt{external variable 1} and \texttt{external variable 3} are selected. For the internal and external covariates LMM, we merge two levels of \texttt{external variable 2} since the point estimates are approximately the same and the 95\% CIs show a large overlap. Hence, in this model we retain the external covariates ${}_{BF} x_{ijkt}$ and \texttt{external variable 2}. Further, based on the results of the multi-type Lasso, \texttt{external variable 1} and \texttt{external variable 3} are removed from the internal and external covariates Tweedie model and ${}_{BF} x_{ijkt}$ is the only remaining external covariate in \eqref{eq:EquationVarSel}. After these adjustments, the internal and external covariate models are refit. \Cref{fig:FinalAllModelLMM,fig:FinalAllModelTweedie} show the estimated coefficients of the external covariates only.

When rounded, the estimated power parameter $\hat{p} = 1.77$ in both the internal covariates only and internal and external covariates Tweedie GLMMs, which corresponds to a claim-size distribution with mode in zero  since $\hat{p} \in (1.5, 2)$ \citep{Jorgensen1994}. This value seems appropriate considering that our data set is characterized by a large amount of $Y_{ijkt} = 0$ (see Section \ref{sec:InternalDataset}).

\begin{figure}[!htbp]
	\centering
	\caption{\label{fig:InspectionREsClass}Random effect estimates of the industries}
	\makebox[\textwidth][c]{\includegraphics[scale = 0.4]{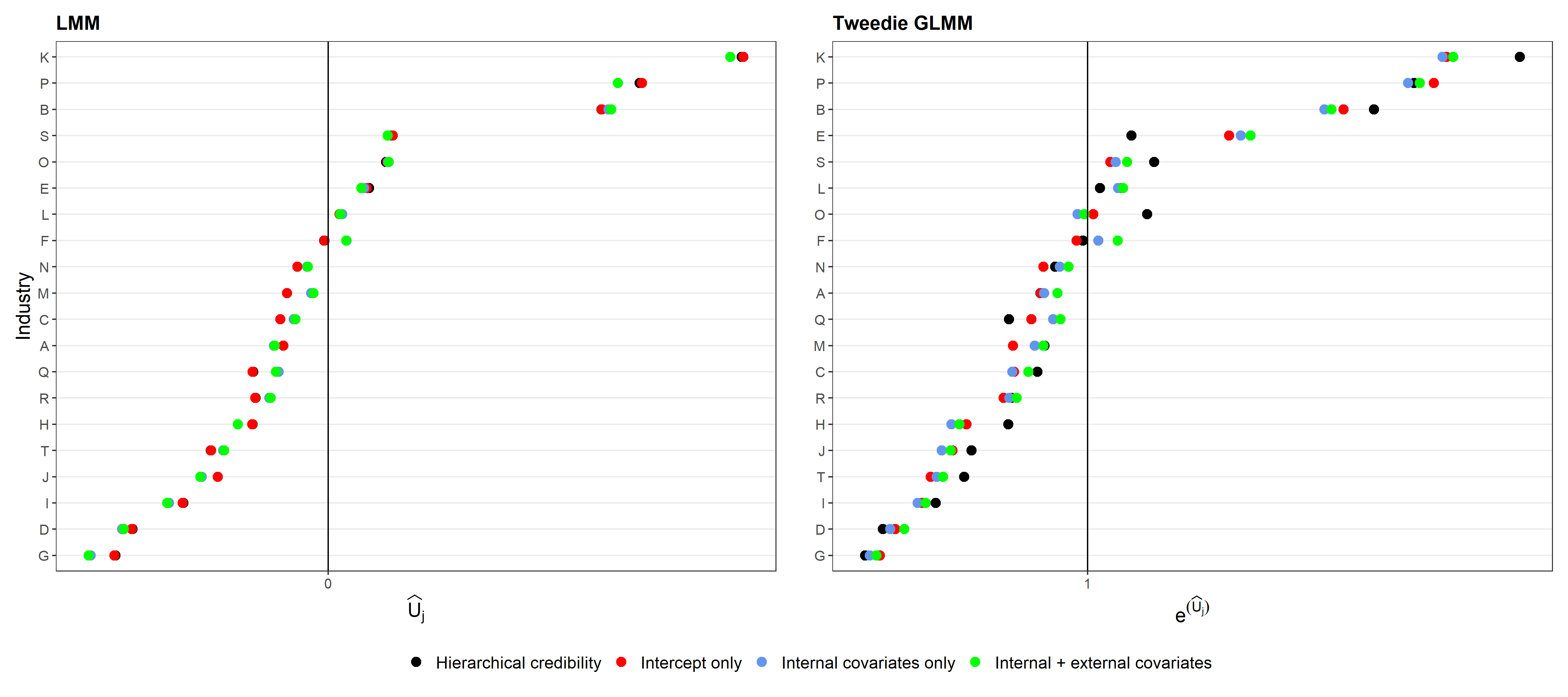}}
\end{figure}

\paragraph*{Examining the random effect estimates.} To examine and compare the random effect estimates across the different estimation methods, we plot the estimates obtained for the industries and branches. \Cref{fig:InspectionREsClass} shows the random effect estimates of the industries for the LMMs and Tweedie GLMMs. In these plots, we add the random effect estimates of the hierarchical credibility model (see Section \ref{sec:2}). For the LMMs, we use the additive Jewell model (see equation \eqref{eq:JewellAdditive}) and for the Tweedie GLMMs, we use the multiplicative Jewell model (see equation \eqref{eq:JewellMultiplicative}). We use the exponent of the random effect estimates of the Tweedie GLMMs to be able to compare these with the estimates resulting from the multiplicative Jewell model. 

The random effect estimates of the LMMs are approximately equal to those of the additive Jewell model. In addition, the random effect estimates of the internal covariates only LMM (blue) and internal and external LMM (green) are nearly identical. This causes the estimates to overlap in the left plot of \Cref{fig:InspectionREsClass}. In contrast, the differences across the different estimation methods are larger for the multiplicative models. Here, we see a large difference between the random effect estimates of the Tweedie GLMMs and the random effect estimates of the multiplicative Jewell model. Comparing the random effect estimates of the industries across the different Tweedie GLMMs, we see that these are slightly higher for the internal and external covariates model compared to estimates of the intercept only and internal covariates only models. In contrast, the random effect estimates of the branches are approximately equal for all Tweedie GLMMs (see Appendix \ref{App:REs}). There are, however, large differences between the random effect estimates of the branches estimated by the multiplicative Jewell model and those estimated by the Tweedie GLMMs. We therefore inspect these estimates in detail for a selected group of branches, but can only give the overall conclusion due to the confidentiality of the data. We observe large contract-specific damage rates $Y_{ijkt}$ as well as high weighted averages $\bar{Y}_{\cdot jk \cdot}$ in branches with large corresponding random effect estimates. Both the estimation method as well as the distributional assumption seem to have an impact on the random effect estimates. The results indicate that the random effect estimates of the Tweedie GLMMs are more in line with the empirical results compared to the random effect estimates of the multiplicative Jewell model.


\subsection{Inspecting the fitted values on the training set}\label{sec:FittedValues}
Next, we examine the fitted values on the company-level training set in more detail. We focus on the (Gaussian and Tweedie) internal covariates only models and we use the hierarchical credibility model as a benchmark. We compute the damage rate for an individual company as 
\begin{equation}\label{eq:DamageRateCompany}
	\begin{aligned}
		Y_{hijkt} = \frac{\mathcal{Z}_{hijkt}}{w_{hijkt}}
	\end{aligned}
\end{equation}
\noindent
where $\mathcal{Z}_{hijkt}$ denotes the capped claim amount of the $h^{th}$ company of tariff class $i$ operating in branch $k$ within industry $j$ at time $t$ and $w_{hijkt}$ is the salary mass. $\widehat{Y}_{hijkt}$ stands for the fitted damage rate. To enable a detailed description of the results whilst preserving the confidentiality of the data, we multiply both $Y_{hijkt}$ and $\widehat{Y}_{hijkt}$ with a constant.


\paragraph*{Balance property.} For insurance applications, it is crucial that the models provide us a reasonable premium volume at portfolio level. Hereto, we examine the balance property \citep{Buhlmann2005,Wuthrich} on the training set. That is,
\begin{equation}
	\begin{aligned}
		\sum_{i, j, k, t} w_{ijkt} \ Y_{ijkt} &= \sum_{i, j, k, t} w_{ijkt} \ \widehat{Y}_{ijkt}\\
	\end{aligned}
\end{equation}
where $i$ serves as an index for the tariff class. GLMs fulfill the balance property when we use the canonical link (see \citet{Wuthrich}). For LMMs and hence, the hierarchical credibility model this property also holds. Conversely, most GLMMs do not have this property. To regain the balance property, we introduce a quantity $\alpha$ 
\begin{equation}
	\begin{aligned}
		\alpha &= \frac{\sum_{i, j, k, t} w_{ijkt} \ Y_{ijkt}}{\sum_{i, j, k, t} w_{ijkt} \ \widehat{Y}_{ijkt}}\\
	\end{aligned}
\end{equation}
which quantifies the deviation of the total predicted damage from the total observed damage. In case of the log link, we can then use $\alpha$ to update the intercept to $\hat{\mu} + \log(\alpha)$ to regain the balance property. We therefore update the intercept for all Tweedie GLMMs, at the level of the training data, and calculate the fitted values using the updated intercept.

\paragraph*{Company-specific covariate levels.} The internal covariates only models contain the first two internal variables and \texttt{two-digit postal code binned}. For each tariff class $i$, we compute the empirical weighted average of the damage rates $\bar{Y}_{\cdot i \cdot \cdot \cdot}$ and weighted average of the predictions $\bar{\widehat{Y}}_{\cdot i \cdot \cdot \cdot}$ using
\begin{equation}
	\begin{aligned}
		\bar{Y}_{\cdot i \cdot \cdot \cdot} = \frac{\sum_{h, j, k, t} w_{hijkt} Y_{hijkt}}{\sum_{h, j, k, t} w_{hijkt}} \ \ \text{and} \ \ \bar{\widehat{Y}}_{\cdot i \cdot \cdot \cdot} = \frac{\sum_{h, j, k, t} w_{hijkt} \widehat{Y}_{hijkt}}{\sum_{h, j, k, t} w_{hijkt}}.
	\end{aligned}
\end{equation}
\noindent

\begin{figure}[!htbp]
	\centering
	\caption{\label{fig:InspectionPredictionsCovariates}Distribution and weighted averages of $Y_{hijkt}$ and $\widehat{Y}_{hijkt}$ for a selected set of tariff classes. Both $Y_{hijkt}$ and $\widehat{Y}_{hijkt}$ are multiplied with a constant to preserve the confidentiality of the data.}
	\makebox[\textwidth][c]{\includegraphics[width = \textwidth]{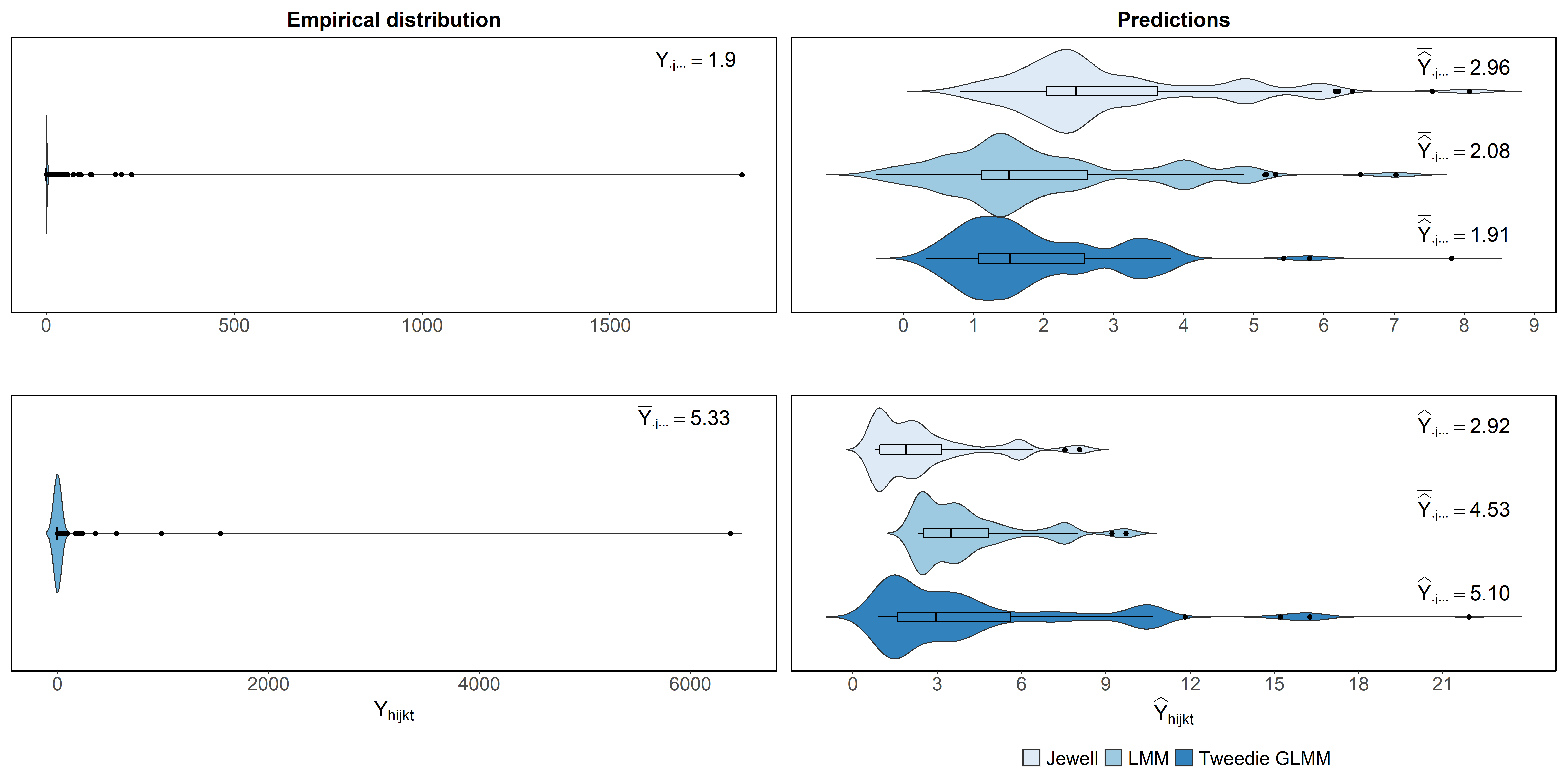}}
\end{figure}

\Cref{fig:InspectionPredictionsCovariates} depicts the results for two different tariff classes. The plots on the left show the empirical distribution of the $Y_{hijkt}$'s together with the $\bar{Y}_{\cdot i \cdot \cdot \cdot}$ . The plots on the right show the distribution of the $\widehat{Y}_{hijkt}$'s and the $\bar{\widehat{Y}}_{\cdot i \cdot \cdot \cdot}$ of the different models. For the majority of the tariff classes, the predictions of the Tweedie model most closely correspond with what we observe empirically. Overall, we observe that, as the range of the $Y_{hijkt}$'s increases, the range of the $\widehat{Y}_{hijkt}$'s increases correspondingly. The predictions are centered at $\bar{Y}_{\cdot i \cdot \cdot \cdot}$ and $\bar{\widehat{Y}}_{\cdot i \cdot \cdot \cdot}$ is approximately equal to $\bar{Y}_{\cdot i \cdot \cdot \cdot}$. In comparison, for the LMM we have negative $\widehat{Y}_{ijkt}$'s and the predictions show a larger deviation from what we observe in the data.

\paragraph*{Hierarchical MLF levels.} To inspect the predictions at the different hierarchical MLF levels, we split the company-level training set using the hierarchical MLF. We compute the empirical weighted average of the damage rates $\bar{Y}_{\cdot \cdot jk \cdot}$ and weighted average of the predictions $\bar{\widehat{Y}}_{\cdot \cdot jk \cdot}$ using
\begin{equation}
	\begin{aligned}
		\bar{Y}_{\cdot \cdot jk \cdot} = \frac{\sum_{h, i, t} w_{hijkt} Y_{hijkt}}{\sum_{h, i, t} w_{hijkt}} \ \ \text{and} \ \ \bar{\widehat{Y}}_{\cdot \cdot jk \cdot} = \frac{\sum_{h, i, t} w_{hijkt} \widehat{Y}_{hijkt}}{\sum_{h, i, t} w_{hijkt}}.
	\end{aligned}
\end{equation}
\noindent

\begin{figure}[!htbp]
	\centering
	\caption{\label{fig:InspectionPredictionsMLF}The distribution and weighted averages of $Y_{hijkt}$ and $\widehat{Y}_{hijkt}$ for branch D4 in industry D and for branch P2 in industry P are shown on the left. The bar plots and map on the right depict the composition of the covariate levels in these branches. Both $Y_{hijkt}$ and $\widehat{Y}_{hijkt}$ are multiplied with a constant to preserve the confidentiality of the data.}
	\makebox[\textwidth][c]{\includegraphics[width = \textwidth]{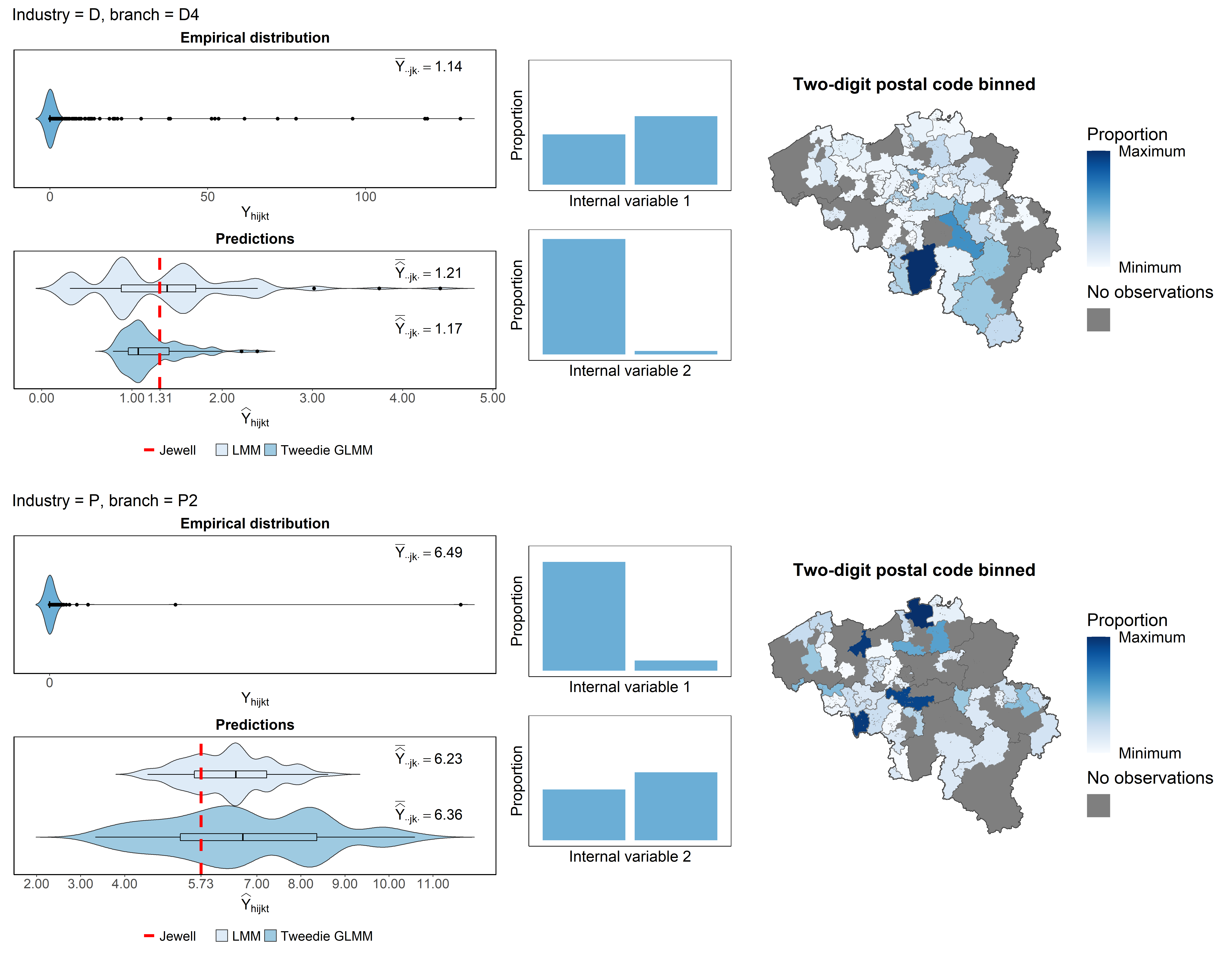}}
\end{figure}

\Cref{fig:InspectionPredictionsMLF} shows the results for branch D4 in industry D (associated with a low random effect of the branch and industry) and for branch P2 in industry P (associated with a high random effect for the branch and industry). As before, the Tweedie GLMM predictions most closely resemble the empirical results in most branches. The range of the $\widehat{Y}_{hijkt}$'s increases as the range of the $Y_{hijkt}$'s increases, $\bar{\widehat{Y}}_{\cdot \cdot jk \cdot}$ is approximately equal to $\bar{Y}_{\cdot \cdot jk \cdot}$ and the predictions are centered at $\bar{Y}_{\cdot \cdot jk \cdot}$. Furthermore, the predominant covariate pattern in a branch determines whether the average prediction of the (G)LMM is lower or higher compared to the prediction of the hierarchical credibility model. Within branch D4, for example, the majority of the observations are categorized into covariate levels that are considered to be less risky relative to the other levels. Consequently, the average prediction of the (G)LMM is lower than the prediction of the hierarchical credibility model. 

\subsection{Assessing the predictive performance}
\begin{figure}[!htbp]
	\centering
	\caption{\label{fig:EmpDistrVal}Empirical distribution of the damage rates $Y_{ijkt}$ of the individual companies in the test set}
	\makebox[\textwidth][c]{\includegraphics[scale = 0.5]{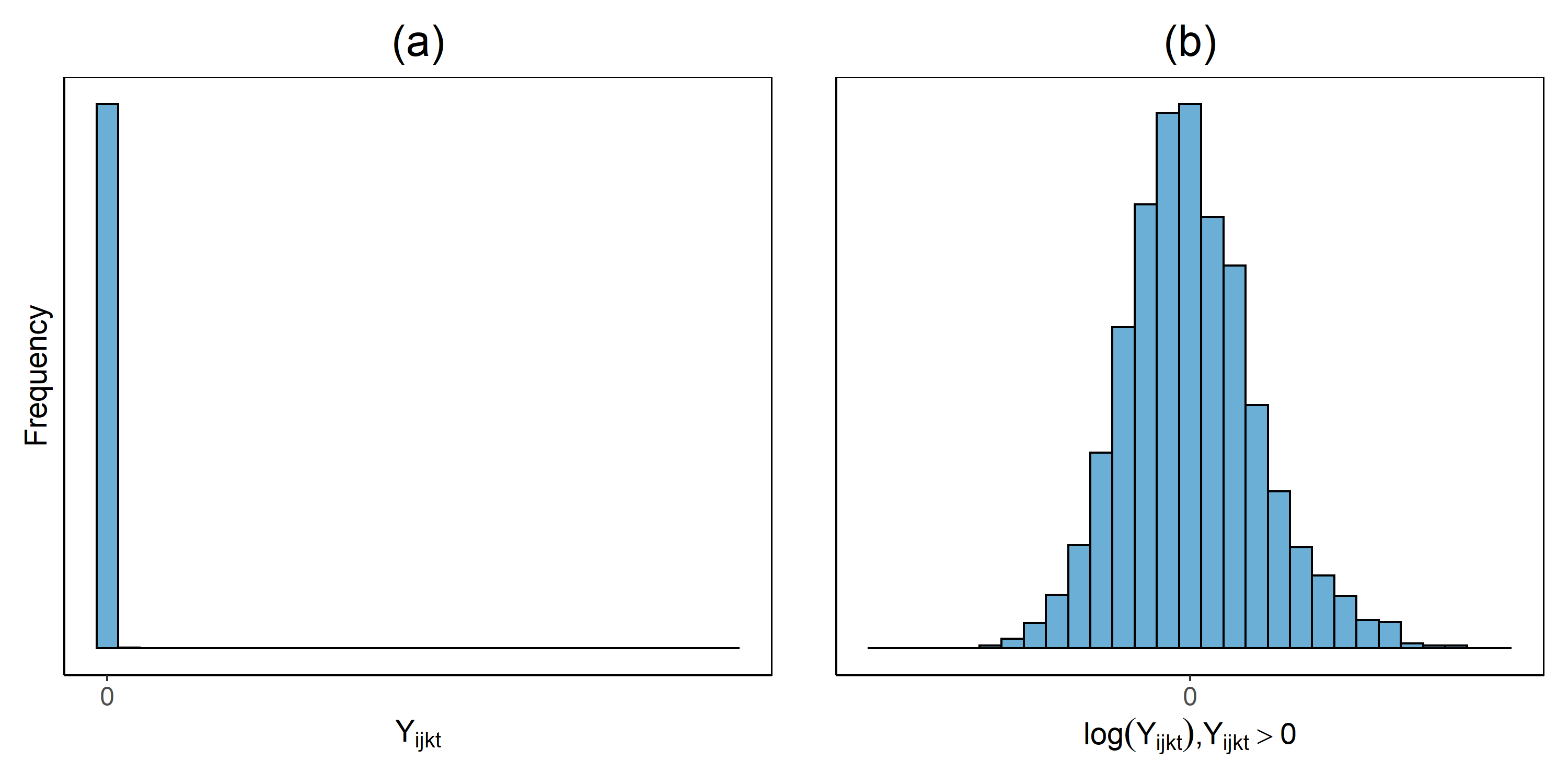}}
\end{figure}

We assess the predictive performance of the pricing model on the test set, which contains damage rates of the individual companies $i$ in the most recent year available. The empirical distribution of the damage rates $Y_{ijkt}$ of the individual companies in the test set is shown in \Cref{fig:EmpDistrVal}. Panel (a) contains all $Y_{ijkt}$'s present in the test set and panel (b) shows the empirical distribution of the log transformed $Y_{ijkt}$ for $Y_{ijkt} > 0$. The empirical distribution of the $Y_{ijkt}$ in the test set is similar to the one observed when using all available data (\Cref{fig:EmpDistr}).

\paragraph*{Performance measures.} To assess the performance of the models, we predict the damage rates in the test set and evaluate the model predictions using the Lorenz curve \citep{Lorenz1905}, Gini-index \citep{Gini1921} and loss ratio. The Lorenz-curve and Gini-index are considered to be appropriate tools to compare competing pricing models \citep{Denuit2019} and assess how well the models are able to differentiate between low- and high-risk companies. Conversely, the loss ratio gives an indication of the overall accuracy of the model predictions. 

The Lorenz curve plots the cumulative percentage of the predicted damage rates against the cumulative proportion of damage rates, with the latter sorted by the predicted damage rates from high to low. An ideal Lorenz curve situates itself in the upper-left corner and indicates that it perfectly distinguishes high-risk companies from low-risk companies. The Gini-index is defined as the ratio of the area between the Lorenz curve and the line of equality ($A$) over the total area between the upper-left corner and the line of equality (= 0.5)

\begin{equation}
	\begin{aligned}
		\text{G} = \frac{A}{0.5} .
	\end{aligned}
\end{equation}
\noindent
For a perfect model, we obtain the maximum theoretical value of $G = 1$. To compute the loss ratio, the total damage on the test set is computed together with the predicted damage by each of the models by transforming the individual predictions $\widehat{Y}_{ijkt}$ as follows (see equation \eqref{eq:KeyRatio})
\begin{equation}
	\begin{aligned}
		\widehat{\mathcal{Z}}_{ijkt} = \widehat{Y}_{ijkt} \ w_{ijkt}.
	\end{aligned}
\end{equation}
\noindent
When we denote the total capped claim amount as $\mathcal{Z}^{tot}_t = \sum_{i, j, k} \mathcal{Z}_{ijkt}$ and the total predicted claim amount as $\widehat{\mathcal{Z}}^{tot}_t = \sum_{i, j, k} \widehat{\mathcal{Z}}_{ijkt}$, the loss ratio is computed as $\mathcal{Z}^{tot}_t / \widehat{\mathcal{Z}}^{tot}_t$. Next to these performance measures, we also inspect the difference in technical premium between the (G)LMMs and the hierarchical credibility model by calculating the relative difference $R_{ijkt}$
\begin{equation}
	\begin{aligned}
		R_{ijkt} = \frac{\widehat{Y}_{ijkt}^{M} - \widehat{Y}_{ijkt}^{J}}{\widehat{Y}_{ijkt}^{J}}
	\end{aligned}
\end{equation}
\noindent
where $\widehat{Y}_{ijkt}^{M}$ denotes the predicted pure premium by the (G)LMMs and $\widehat{Y}_{ijkt}^{J}$ the predicted pure premium by the hierarchical credibility model, which serves as the benchmark model. This allows us to identify both overpriced and underpriced policies. Compared to the hierarchically credibility model, policies are currently overpriced when $R_{ijkt} < 0$ and can potentially be lost to competitors. Conversely, $R_{ijkt} > 0$ indicates that the policy is underpriced compared to the hierarchical credibility model and this necessitates appropriate loss control measures to prevent future financial losses.

\paragraph*{Out-of-sample performance.} Table \ref{tab:PredPerfOOS} summarizes the out-of-sample performance of the models on the test set. For both the LMM and Tweedie GLMM, the Gini-index increases when company-specific risk factors are included. Consequently, by adding company-specific risk factors we are better able to distinguish high- from low-risk companies compared to when we do not include these in the model. Further, in both the LMMs and Tweedie GLMMs the model performance decreases when we add external covariates. In addition, the loss ratio of the internal and external covariates LMM is higher than the loss ratio of the internal covariates only LMM. When the external covariate ${}_{BF} x_{ijkt}$ is added to the internal covariates only Tweedie GLMM, the loss ratio shows a slight improvement. Comparing the internal covariates only LMM and internal covariates only Tweedie GLMM, we see that the predictive performance of the Tweedie model is better. The Gini-index is higher and the loss ratio is closer to one, indicating that the Tweedie GLMM is better able to differentiate between low- and high-risk companies and results in a more accurate estimation of the total damage. In addition, the loss ratio of the internal covariates only Tweedie GLMM is lower than the loss ratio of the hierarchical credibility model.

\setlength{\tabcolsep}{5pt}
\begin{table}[ht]
	\centering
	\caption{Comparison predictive performance on the test set} 
	\label{tab:PredPerfOOS}
	\makebox[\textwidth][c]{\begin{tabular}{lllcHHcHHHH}
			\hline
			Model & Distribution & Variable set & Gini-index & Total damage & Total predicted damage & Loss ratio & Overall calibration & wMSE & logLik & AIC \\ 
			\hline
			Jewell &  &  & 0.592 & 63 824 906 & 63 269 696 & 1.009 & 1.433 & 621.064 &  &  \\ 
			LMM & Gaussian & Intercept only & 0.591 & 63 824 906 & 63 267 393 & 1.009 & 1.436 & 621.068 &  &  \\ 
			&  & Internal covariates & 0.653 & 63 824 906 & 62 978 495 & 1.013 & 1.430 & 620.807 &  &  \\ 
			&  & Internal + external covariates& 0.644 & 63 824 906 & 61 828 292 & 1.032 & 1.411 & 620.797 &  &  \\
			GLMM & Tweedie & Intercept only & 0.607 & 63 824 906 & 63 223 044 & 1.010 & 1.221 & 621.007 &  &  \\ 
			&  & Internal covariates & 0.660 & 63 824 906 & 63 402 798 & 1.007 & 1.504 & 620.988 &  &  \\ 
			&  & Internal + external covariates& 0.650 & 63 824 906 & 63 444 780 & 1.006 & 1.496 & 620.969 &  &  \\
			\hline
	\end{tabular}}
\end{table}
\setlength{\tabcolsep}{6pt}

\begin{figure}[!ht]
	\centering
	\caption{\label{fig:LorenzCurve}Lorenz curves}
	\makebox[\textwidth][c]{\includegraphics[scale = 0.35]{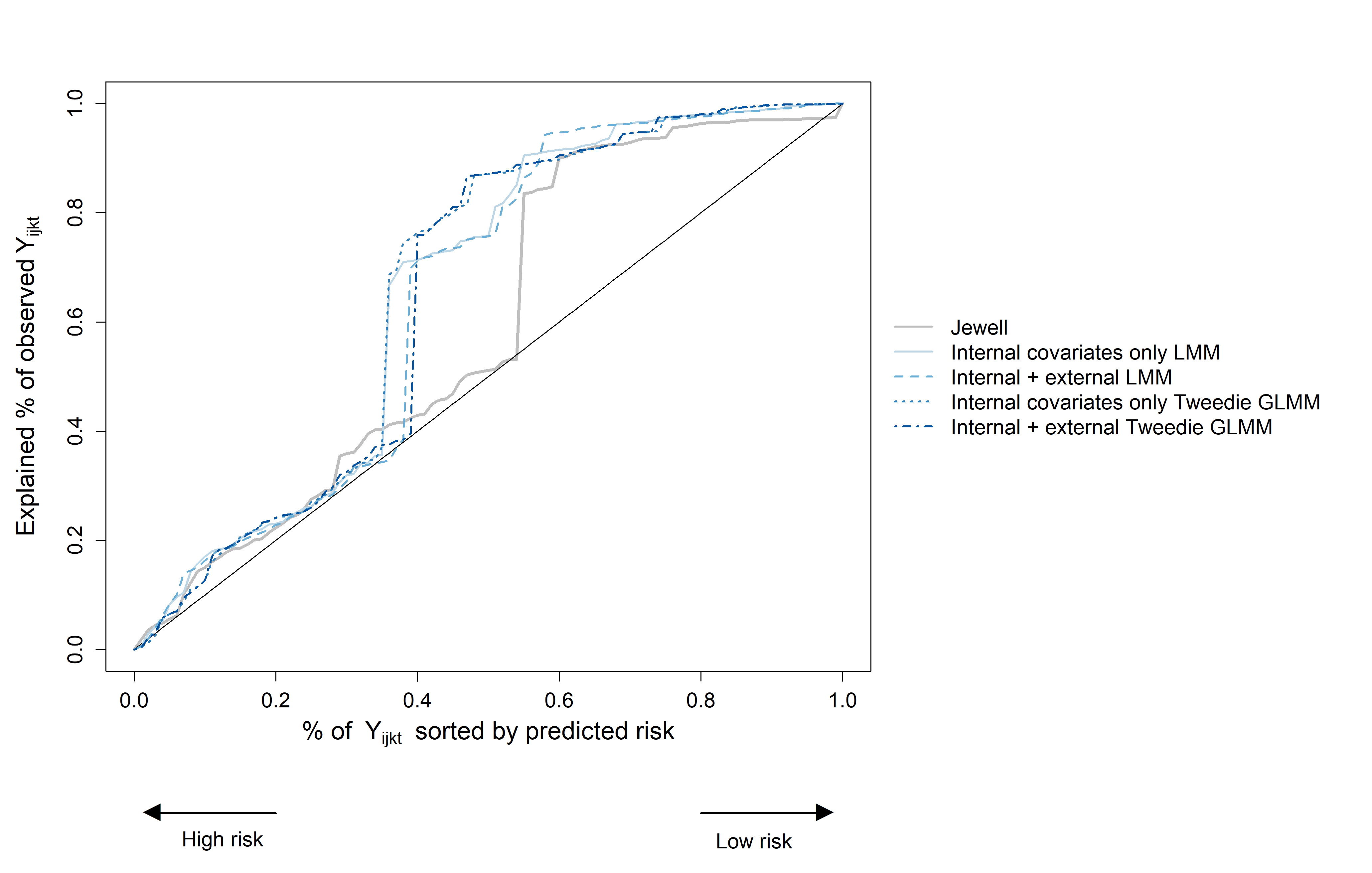}}
\end{figure}

Figure \ref{fig:LorenzCurve} shows the Lorenz curves of the different models. The hierarchical credibility model has the lowest performance and the internal covariates only Tweedie GLMM the best performance. For all models the Lorenz curve is close to the diagonal line for observations that are considered to be high-risk. This indicates that the models experience difficulties with accurately ordering companies characterized by high $\widehat{Y}_{ijkt}$'s. Conversely, when the predicted risk decreases, the ordering of the companies gets more accurate as the Lorenz curves are further removed from the diagonal line. Consequently, all models are better able to differentiate high- from low-risk companies that have medium to low predicted damage rates. 

\begin{figure}[!ht]
	\centering
	\caption{\label{fig:RelPremium}Relative premium differences on the test set}
	\makebox[\textwidth][c]{\includegraphics[scale = 0.35]{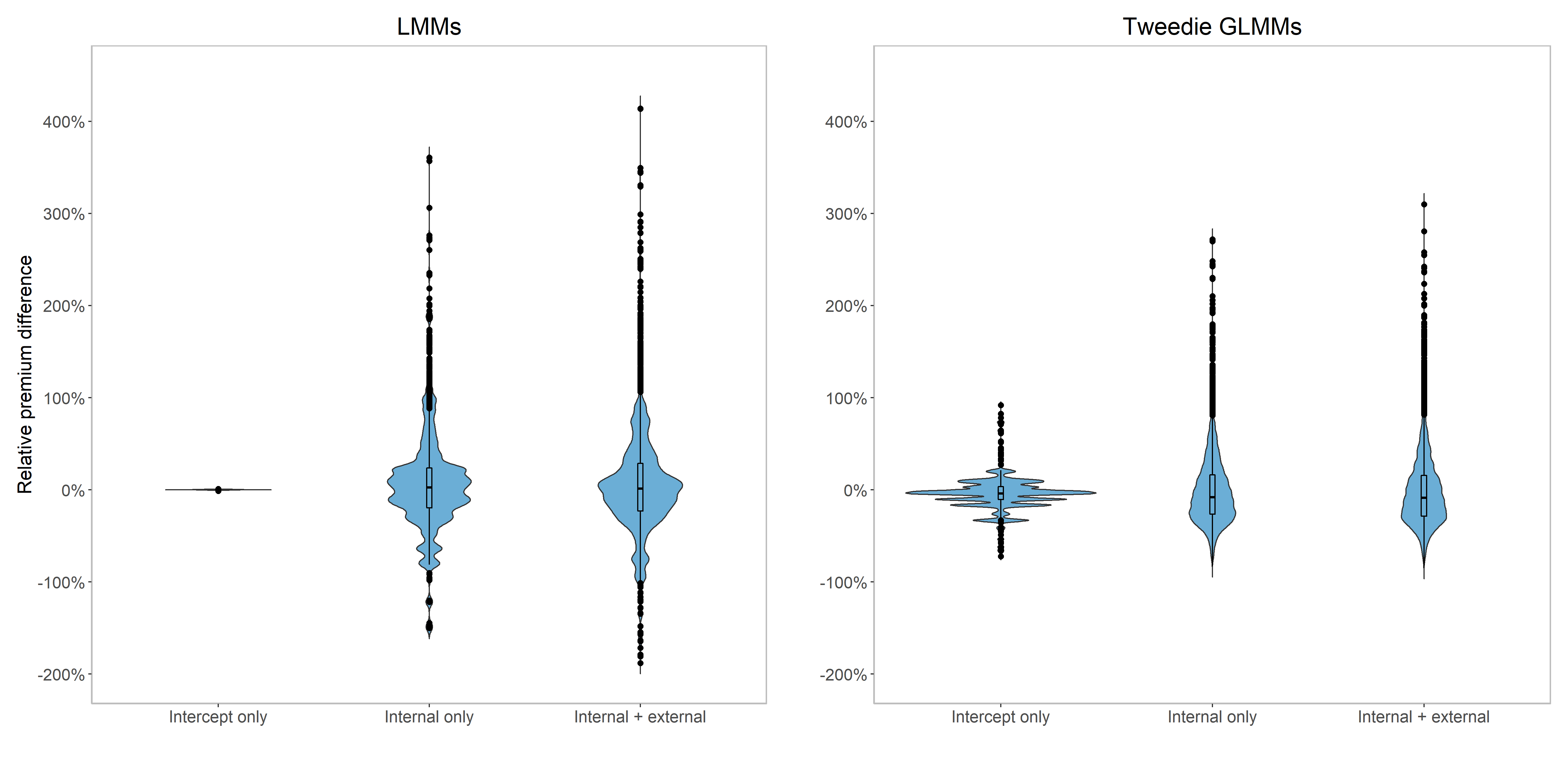}}
\end{figure}

\Cref{fig:RelPremium} depicts the relative premium differences. The difference is negligible when using the intercept-only LMM, which is due to the equivalence between the intercept-only LMM and Jewell model. Larger differences are seen in the $\widehat{Y}_{ijkt}$'s when using the Tweedie intercept-only GLMM which is caused by larger differences in the random effect estimates. When adding company-specific risk factors to the LMM and Tweedie GLMM, the majority of the companies see a decrease in the expected pure premium. Here, the density is larger for $R_{ijkt} < 0$ indicating that $\widehat{Y}_{ijkt}^{M} < \widehat{Y}_{ijkt}^{J}$. In addition, for most companies that see an increase in the expected pure premium when company-specific risk factors are added, this will be a 50\% increase of the pure premium at most and this will be more than 50\% for only a few companies. 

\section{Discussion}\label{sec:4}
In this paper, we show how random effects models can be used to construct a data-driven insurance pricing model when working with hierarchically structured data supplemented by internal and external contract-specific risk factors. We examine several random effects models previously proposed within the actuarial literature, such as the hierarchical credibility model of \citet{JewellModel}, the combination of a GLM with the hierarchical credibility model \citep{Ohlsson2008} and mixed models. We examine and compare the performance of these random effects models using a workers' compensation insurance portfolio from a Belgian insurer. In addition, we assess the effect of the distributional assumption of the response as well as the added value of contract-specific risk factors derived from an external data source.

The random effects specification allows us to efficiently estimate and quantify the effect of the different hierarchical MLF levels. Further, incorporating contract-specific information in the model results in an improved predictive performance. With regard to the estimation methods, we find that Ohlsson's iterative GLMC algorithm is ideal in combination with (exhaustive) variable selection methods. Its simplicity and computational efficiency allows for a quick estimation of the parameters. In addition, the parameter estimates can be used as starting values when fitting GLMMs. The GLMMs are computationally heavy and are prone to convergence issues. Providing appropriate starting values drastically speeds up the GLMM algorithm and frequently helps to overcome convergence issues. Given their well-developed statistical framework, which allows for statistical inference, the GLMMs are well suited to examine the model and can be used as a final estimation step to obtain accurate estimates. With regard to the distributional assumptions, the results indicate that the Gaussian distribution is not ideal in combination with company-specific covariates. Due to the presence of zero valued claims and the symmetric nature of the Gaussian distribution, some companies obtain a negative predicted damage rate. Conversely, the Tweedie distribution is especially suited for modeling and predicting damage rates. As previously stated by \citet{Jorgensen1994}, the Tweedie distribution handles zero valued observations in a natural, satisfactory way. Moreover, compared to the LMMs, the Tweedie GLMMs are better able to differentiate between low- and high-risk companies and result in a more accurate estimation of the total claim amount in the test set. In addition, including company-specific covariates allows for more and better differentiation between companies and the Tweedie model is possibly better able to detect groups characterized by large damage rates. Adding external company-specific covariates to the internal covariates only model, however, did not result in significant improvements.

The absence of an added value of incorporating external data may be caused by limiting ourselves to one specific external data source. Future research can examine whether this generalizes to other data sets and examine the potential predictive value of other external data sources.\ignore{In addition, the results indicate that our not missing at random assumption for the missing values is questionable. More research is warranted on which methods are feasible and appropriate to handle missings when the missing at random assumption is violated.} Further, we limit ourselves to regression-type random effects models. The theoretical framework can be extended to include random effects machine-learning methods as well, such as the RE-EM tree of \citet{REtrees}. Given the promising results of machine-learning methods within actuarial applications, it may prove to be worthwhile to examine whether this generalizes to hierarchically structured data as well.

\addcontentsline{toc}{chapter}{Bibliography}
\bibliographystyle{agsmAdj}
\renewcommand\harvardurl{}
\bibliography{ReferencesSAJ}

@book{Denuit2019Book,
series = {Springer actuarial lecture notes},
publisher = {Springer International Publishing AG},
booktitle = {<h>Effective</h> <h>Statistical</h> <h>Learning</h> <h>Methods</h> <h>for Actuaries</h> <h>I</h>},
isbn = {303025819X},
year = {2019},
title = {Effective statistical learning methods for actuaries I: GLMs and extensions},
language = {eng},
address = {Cham},
author = {Denuit, Michel and Hainaut, Donatien and Trufin, Julien},
}

@book{GLM,
series = {Monographs on statistics and applied probability 37},
publisher = {Chapman and Hall},
isbn = {0412317605},
year = {1999},
title = {Generalized linear models},
address = {London},
author = {McCullagh, P and Nelder, J. A},
}

@book{Ohlsson,
series = {EAA Lecture Notes},
issn = {3-642-10790-7},
publisher = {Springer},
isbn = {1280391626},
year = {2010},
title = {Non-life insurance pricing with generalized linear models},
language = {eng},
address = {Berlin, Heidelberg},
author={Ohlsson, E. and Johansson, B.},
}

@article{Ohlsson2008,
issn = {0346-1238},
journal = {Scandinavian Actuarial Journal},
pages = {301--314},
volume = {2008},
publisher = {Taylor \& Francis Group},
number = {4},
year = {2008},
title = {Combining generalized linear models and credibility models in practice},
author = {Ohlsson, Esbj{\"o}rn},
}

@techreport{Reacfin,
year = {2017},
author = {Stassen, B. and Denuit, M. and Mahy, S. and Mar{\'e}chal, X. and J. Trufin},
title = {A unified approach for the modelling of rating factors in workers compensation insurance},
type = {White paper by {Reacfin}},
note = {Available at: \url{https://www.reacfin.com/wp-content/uploads/2016/12/170131-Reacfin-White-Paper-A-Unified-Approach-for-the-Modeling-of-Rating-Factors-in-Work-ers\%E2\%80\%99-Compensation-Insurance.pdf}}
}

@article{Henckaerts2020,
issn = {1092-0277},
journal = {North American Actuarial Journal},
pages = {255--285},
volume = {25},
publisher = {Routledge},
number = {2},
year = {2021},
title = {Boosting Insights in Insurance Tariff Plans with Tree-Based Machine Learning Methods},
copyright = {2020 Society of Actuaries 2020},
language = {eng},
author = {Henckaerts, Roel and {C{\^o}t{\'e}}, Marie-Pier and Antonio, Katrien and Verbelen, Roel},
}

@book{Denuit2007,
publisher = {Wiley},
isbn = {9780470026779},
year = {2007},
title = {Actuarial modelling of claim counts: risk classification, credibility and bonus-malus systems},
address = {Chichester},
author = {Denuit, Michel and Mar{\'e}chal, Xavier and Pitrebois, Sandra and Walhin, Jean-François},
}

@book{Frees2014,
  title={Predictive modeling applications in actuarial science: volume 1, predictive modeling techniques},
  author={Frees, E.W. and Derrig, R.A. and Meyers, G.},
  isbn={9781139992312},
  series={International Series on Actuarial Science},
  url={https://books.google.be/books?id=2ENpBAAAQBAJ},
  year={2014},
  publisher={Cambridge University Press},
  address = {New York},
}

@book{Parodi2014,
  title={Pricing in general insurance},
  author={Parodi, P.},
  isbn={9781466581487},
  url={https://books.google.be/books?id=BGnSBQAAQBAJ},
  year={2014},
  publisher={CRC Press},
  address = {New York}
}

@article{Haberman1996,
issn = {0039-0526},
journal = {Journal of the Royal Statistical Society: Series D (The Statistician)},
pages = {407--436},
volume = {45},
number = {4},
year = {1996},
title = {Generalized Linear Models and Actuarial Science},
author = {Haberman, Steven and Renshaw, Arthur E.},
}

@book{Dejong2008,
  title={Generalized linear models for insurance data},
  author={de Jong, P. and Heller, G.Z.},
  isbn={9781139470476},
  series={International Series on Actuarial Science},
  url={https://books.google.be/books?id=DW0syblRyHUC},
  year={2008},
  publisher={Cambridge University Press},
  address = {Cambridge}
}

@article{Frees2015,
author = {Frees, E.W},
year = {2015},
month = {12},
pages = {253-277},
title = {Analytics of Insurance Markets},
volume = {7},
journal = {Annual Review of Financial Economics},
doi = {10.1146/annurev-financial-111914-041815}
}

@article{Frees2008,
issn = {0162-1459},
journal = {Journal of the American Statistical Association},
pages = {1457--1469},
volume = {103},
publisher = {Taylor \& Francis},
number = {484},
year = {2008},
title = {Hierarchical Insurance Claims Modeling},
language = {eng},
author = {Frees, Edward W. and Valdez, Emiliano A.},
}

@article{Antonio2010,
issn = {0515-0361},
journal = {ASTIN Bulletin},
pages = {151--177},
volume = {40},
number = {1},
year = {2010},
title = {A multilevel analysis of intercompany claim counts},
language = {eng},
author = {Antonio, K. and Frees, E.W. and Valdez, E.A.},
}

@book{Molenberghs2005,
series = {Springer Series in Statistics},
publisher = {Springer New York},
isbn = {9780387251448},
year = {2005},
title = {Models for discrete longitudinal data},
language = {eng},
address = {New York},
author = {Molenberghs, Geert and Verbeke, Geert},
}

@article{Lorenz1905,
issn = {1522-5437},
journal = {Publications of the American Statistical Association},
pages = {209--219},
volume = {9},
publisher = {Taylor \& Francis Group},
number = {70},
year = {1905},
title = {Methods of Measuring the Concentration of Wealth},
author = {Lorenz, M. O.},
}

@article{Gini1921,
issn = {00130133},
journal = {The Economic Journal},
pages = {124--126},
volume = {31},
publisher = {MacMillan and Co. Limited},
number = {121},
year = {1921},
title = {Measurement of Inequality of Incomes},
language = {eng},
author = {Gini, Corrado},
}

@article{REvariance2,
issn = {0006-341X},
journal = {Biometrics},
pages = {270--279},
volume = {67},
publisher = {Blackwell Publishing Inc},
number = {1},
year = {2011},
title = {Prediction of Random Effects in Linear and Generalized Linear Models under Model Misspecification},
address = {Malden, USA},
author = {McCulloch, Charles E. and Neuhaus, John M.},
}

@misc{ConvergenceWarnings,
  title = {{GLMM FAQ}},
  howpublished = {\url{https://bbolker.github.io/mixedmodels-misc/glmmFAQ.html#convergence-warnings}},
  note = {Viewed 07 November 2022},
  author = {Bolker, Ben and others},
  year = {2022},
  month = {10}
}

@Manual{Rsoftware,
    title = {R: A Language and Environment for Statistical Computing},
    author = {{R Core Team}},
    organization = {R Foundation for Statistical Computing},
    address = {Vienna, Austria},
    year = {2019},
    url = {https://www.R-project.org/},
  }

@article{FreesIME,
issn = {0167-6687},
journal = {Insurance Mathematics and Economics},
pages = {229--247},
volume = {24},
publisher = {Elsevier B.V.},
number = {3},
year = {1999},
title = {A longitudinal data analysis interpretation of credibility models},
language = {eng},
author = {Frees, Edward W. and Young, Virginia R. and Luo, Yu},
}

@article{AIC,
issn = {0018-9286},
journal = {IEEE Transactions on Automatic Control},
pages = {716--723},
volume = {19},
publisher = {IEEE},
number = {6},
year = {1974},
title = {A new look at the statistical model identification},
language = {eng},
author = {Akaike, H},
}

@Manual{smurf,
    title = {smurf: Sparse Multi-Type Regularized Feature Modeling},
    author = {Tom Reynkens and Sander Devriendt and Katrien Antonio},
    year = {2018},
    note = {R package version 1.0.0},
    url = {https://CRAN.R-project.org/package=smurf},
  }

@article{OhlssonJewell,
  title={Simplified estimation of structure parameters in hierarchical credibility},
  author = {Ohlsson, Esbj{\"o}rn},
  year={2005},
  note = {Presented at the Zurich ASTIN Colloquium},
  url = {\url{http://www.actuaries.org/ASTIN/Colloquia/Zurich/Ohlsson.pdf}}, 
}

@book{JewellModel,
series = {International Institute for applied systems analysis. Research memoranda RM-75-24},
publisher = {IIASA},
year = {1975},
title = {The use of collateral data in credibility theory: a hierarchical model},
address = {Laxenburg},
author = {Jewell, William S},
}

@book{Dannenburg,
  title={Practical actuarial credibility models},
  author={D. R. Dannenburg and Rob Kaas and Marc J. Goovaerts},
  publisher = {IAE (Institute of Actuarial Science and Econometrics of the University of Amsterdam)},
  address = {Amsterdam},
  year={1996},
}

@article{Subset1,
issn = {0006-3444},
journal = {Biometrika},
pages = {357--366},
volume = {54},
publisher = {Biometrika Office, University College, London},
number = {3},
year = {1967},
title = {The Discarding of Variables in Multivariate Analysis},
language = {eng},
address = {England},
author = {E. M. L. Beale and M. G. Kendall and D. W. Mann},
}

@article{Subset2,
issn = {0040-1706},
journal = {Technometrics},
pages = {531--540},
volume = {9},
publisher = {Taylor \& Francis Group},
number = {4},
year = {1967},
title = {Selection of the Best Subset in Regression Analysis},
copyright = {Copyright Taylor \&amp; Francis Group, LLC 1967},
language = {eng},
author = {Hocking, R. R and Leslie, R. N},
}

@article{Henckaerts2018,
issn = {0346-1238},
journal = {Scandinavian Actuarial Journal},
pages = {681--705},
volume = {2018},
publisher = {Taylor \& Francis},
number = {8},
year = {2018},
title = {A data driven binning strategy for the construction of insurance tariff classes},
copyright = {2018 Informa UK Limited, trading as Taylor \&amp; Francis Group 2018},
language = {eng},
author = {Henckaerts, Roel and Antonio, Katrien and Clijsters, Maxime and Verbelen, Roel},
}

@article{BinningPC,
issn = {2073-4859},
journal = {The R journal},
pages = {29--33},
volume = {3},
publisher = {The R Foundation},
number = {2},
year = {2011},
title = {Ckmeans.1d.dp: Optimal k-means Clustering in One Dimension by Dynamic Programming},
language = {eng},
address = {United States},
author = {Wang, Haizhou and Song, Mingzhou},
}

@book{StatExtremes,
series = {Wiley series in probability and statistics},
publisher = {Wiley},
isbn = {0471976474},
year = {2005},
title = {Statistics of extremes: theory and applications},
address = {Chichester},
author = {Beirlant, Jan and Goegebeur, Yuri and Segers, Johan and Teugels, Jozef L},	
}

@article{Adalet2018,
issn = {0266-4658},
journal = {Economic policy},
pages = {685--736},
volume = {33},
number = {96},
year = {2018},
title = {The walking dead? {Zombie} firms and productivity performance in {OECD} countries},
language = {eng},
author = {McGowan, M{\"u}ge Adalet and Andrews, Dan and Millot, Valentine},
}

@article{Vanacker2008,
issn = {1573-0913},
journal = {Small Business Economics},
pages = {53--69},
volume = {35},
publisher = {Springer Science and Business Media LLC},
number = {1},
year = {2008},
title = {Pecking order and debt capacity considerations for high-growth companies seeking financing},
copyright = {2010 Springer},
language = {eng},
address = {Boston},
author = {Vanacker, Tom R and Manigart, Sophie},
}

@article{Denuit2019,
issn = {0167-6687},
journal = {Insurance, Mathematics \& Economics},
pages = {128--139},
volume = {89},
publisher = {Elsevier B.V},
year = {2019},
title = {Model selection based on {Lorenz} and concentration curves, {Gini} indices and convex order},
copyright = {2019 Elsevier B.V.},
language = {eng},
author = {Denuit, Michel and Sznajder, Dominik and Trufin, Julien},
}

@article{AntonioGLMM,
series = {Insurance: Mathematics and Economics},
issn = {0167-6687},
journal = {Insurance, Mathematics \& Economics},
pages = {58--76},
volume = {40},
publisher = {Elsevier B.V},
number = {1},
year = {2007},
title = {Actuarial statistics with generalized linear mixed models},
copyright = {2006 Elsevier Ltd},
language = {eng},
author = {Antonio, Katrien and Beirlant, Jan},
}

@article{SmythTweedie2002,
issn = {0515-0361},
journal = {ASTIN Bulletin},
pages = {143--157},
volume = {32},
publisher = {Cambridge University Press},
number = {1},
year = {2002},
title = {Fitting {Tweedie}'s Compound {Poisson} Model to Insurance Claims Data: Dispersion Modelling},
copyright = {Copyright © International Actuarial Association 2002},
language = {eng},
address = {Cambridge, UK},
author = {Smyth, Gordon K and J{\o}rgensen, Bent},
}

@article{Jorgensen1994,
author = { Bent   J{\o}rgensen  and  Marta  C.   Paes De Souza },
title = {Fitting Tweedie's compound poisson model to insurance claims data},
journal = {Scandinavian Actuarial Journal},
volume = {1994},
number = {1},
pages = {69-93},
year  = {1994},
publisher = {Taylor & Francis},
doi = {10.1080/03461238.1994.10413930},
}

@book{RegrModelingActuarial,
series = {International series on actuarial science},
publisher = {Cambridge University Press},
isbn = {0521135966},
year = {2010},
title = {Regression modeling with actuarial and financial applications},
author = {Frees, Edward W},
address = {New York},
}

@article{NormalREs,
issn = {0006-341X},
journal = {Biometrics},
pages = {63--71},
volume = {73},
publisher = {Wiley Subscription Services, Inc},
number = {1},
year = {2017},
title = {Diagnosing misspecification of the random-effects distribution in mixed models},
copyright = {2016, The International Biometric Society},
language = {eng ; fre},
address = {United States},
author = {Drikvandi, Reza and Verbeke, Geert and Molenberghs, Geert},
}

@article{Wuthrich,
issn = {2190-9733},
journal = {European Actuarial Journal},
pages = {179--202},
volume = {10},
publisher = {Springer Nature B.V},
number = {1},
year = {2020},
title = {Bias regularization in neural network models for general insurance pricing},
copyright = {EAJ Association 2019.},
language = {eng},
address = {Heidelberg},
author = {W{\"u}thrich, Mario V},
}

@article{Yang2018,
issn = {0735-0015},
journal = {Journal of Business \& Economic Statistics},
pages = {456--470},
volume = {36},
publisher = {Taylor \& Francis},
number = {3},
year = {2018},
title = {Insurance Premium Prediction via Gradient Tree-Boosted {Tweedie Compound Poisson} Models},
copyright = {2018 American Statistical Association 2018},
language = {eng},
address = {ALEXANDRIA},
author = {Yang, Yi and Qian, Wei and Zou, Hui},
}

@article{Bennett2001,
issn = {1326-0200},
journal = {Australian and New Zealand Journal of Public Health},
pages = {464--469},
volume = {25},
publisher = {Blackwell Publishing Ltd},
number = {5},
year = {2001},
title = {How can {I} deal with missing data in my study},
edition = {Revision requested: November 2000 Accepted: May 2001},
copyright = {Copyright 2018 Elsevier B.V., All rights reserved.},
language = {eng},
address = {Oxford, UK},
author = {Bennett, Derrick A},
}

@article{Saefken2014,
issn = {1935-7524},
journal = {Electronic Journal of Statistics},
volume = {8},
number = {1},
year = {2014},
title = {A unifying approach to the estimation of the conditional Akaike information in generalized linear mixed models},
language = {eng},
author = {Saefken, Benjamin and Kneib, Thomas and van Waveren, Clara-Sophie and Greven, Sonja},
}

@article{Blier2021,
issn = {2227-9091},
journal = {Risks},
pages = {4},
volume = {9},
publisher = {MDPI AG},
number = {4},
year = {2021},
title = {Machine Learning in {P\&C} Insurance: A Review for Pricing and Reserving},
language = {eng},
author = {Christopher Blier-Wong and H{\'e}l{\`e}ne Cossette and Luc Lamontagne and Etienne Marceau},
}

@article{Skrondal2009,
series = {Journal of the Royal Statistical Society Series A},
issn = {0964-1998},
journal = {Journal of the Royal Statistical Society. Series A, Statistics in society},
pages = {659--687},
volume = {172},
publisher = {Blackwell Publishing Ltd},
number = {3},
year = {2009},
title = {Prediction in multilevel generalized linear models},
edition = {Received February 2008. Final revision October 2008},
copyright = {Copyright 2009 The Royal Statistical Society and Blackwell Publishing Ltd.},
language = {eng},
address = {Oxford, UK},
author = {Skrondal, Anders and Rabe-Hesketh, Sophia},
}

@article{Micci2001,
issn = {1931-0145},
journal = {SIGKDD explorations},
pages = {27--32},
volume = {3},
publisher = {ACM},
number = {1},
year = {2001},
title = {A preprocessing scheme for high-cardinality categorical attributes in classification and prediction problems},
language = {eng},
author = {Micci-Barreca, Daniele},
}

@book{NACE,
author = {{European Commission and Eurostat}},
title = {{NACE Rev. 2}: statistical classification of economic activities in the {European} Community},
publisher = {Publications Office},
year = {2017},
}

@article{REtrees,
issn = {0885-6125},
journal = {Machine learning},
pages = {169--207},
volume = {86},
publisher = {Springer US},
number = {2},
year = {2012},
title = {{RE-EM} trees: a data mining approach for longitudinal and clustered data},
copyright = {The Author(s) 2011},
language = {eng},
address = {Boston},
author = {Sela, Rebecca J and Simonoff, Jeffrey S},
}

@article{Tuerlinckx2006,
issn = {0007-1102},
journal = {British Journal of Mathematical \& Statistical Psychology},
pages = {225--255},
volume = {59},
publisher = {Blackwell Publishing Ltd},
number = {2},
year = {2006},
title = {Statistical inference in generalized linear mixed models: A review},
edition = {Received 17 June 2003; revised version received 31 August 2005; Accepted 5 October 2005},
copyright = {2006 The British Psychological Society},
language = {eng},
address = {Oxford, UK},
author = {Tuerlinckx, Francis and Rijmen, Frank and Verbeke, Geert and De Boeck, Paul},
}

@misc{BelFirst,
author = {{Bureau Van Dijk}},
publisher = {Bureau Van Dijk},
year = {2020},
title = {Bel-first: financial reports and statistics on {Belgian} and {Luxembourg} companies},
howpublished = {https://belfirst.bvdinfo.com},
note = {Viewed 12 March 2021},
}

@book{Buhlmann2005,
publisher = {Springer},
isbn = {3540257535},
year = {2006},
title = {A course in credibility theory and its applications},
language = {eng},
address = {Berlin/Heidelberg},
author = {B{\"u}hlmann, Hans and Gisler, Alois},
}

@article{Delong2021,
issn = {2190-9733},
journal = {European Actuarial Journal},
pages = {185--226},
volume = {11},
publisher = {Springer Nature B.V},
number = {1},
year = {2021},
title = {Making {Tweedie}'s compound {Poisson} model more accessible},
copyright = {The Author(s) 2021. This work is published under http://creativecommons.org/licenses/by/4.0/ (the “License”). Notwithstanding the ProQuest Terms and Conditions, you may use this content in accordance with the terms of the License.},
language = {eng},
address = {Heidelberg},
author = {Delong, {\L}ukasz and Lindholm, Mathias and W{\"u}thrich, Mario V},
}

@misc{WCdef,
  title = {{Solvency II single rulebook}},
  howpublished = {\url{https://www.eiopa.europa.eu/rulebook/solvency-ii/article-6339_en?source=search}},
  note = {Viewed 2021-11-26},
  author = {{European Insurance and Occupational Pensions Authority}},
  year = {2020},
  month = {07}
}

@article{cplm,
author = {Zhang, Yanwei},
year = {2013},
month = {11},
pages = {},
title = {Likelihood-based and {Bayesian} methods for {Tweedie Compound Poisson} linear mixed models},
volume = {23},
journal = {Statistics and Computing},
doi = {10.1007/s11222-012-9343-7}
}

@Article{lme4,
  title = {Fitting Linear Mixed-Effects Models Using {lme4}},
  author = {Douglas Bates and Martin M{\"a}chler and Ben Bolker and Steve Walker},
  journal = {Journal of Statistical Software},
  year = {2015},
  volume = {67},
  number = {1},
  pages = {1--48},
  doi = {10.18637/jss.v067.i01},
}

@book{Burnham2007,
  title={Model Selection and Multimodel Inference: A Practical Information-Theoretic Approach},
  author={Burnham, K.P. and Anderson, D.R.},
  isbn={9780387224565},
  lccn={2001057677},
  year={2002},
  publisher = {Springer},
  address = {New York},
  edition = {2nd ed.},
}

@article{Colbrook2022,
author = {Matthew J. Colbrook  and Vegard Antun  and Anders C. Hansen },
title = {The difficulty of computing stable and accurate neural networks: On the barriers of deep learning and Smale's 18th problem},
journal = {Proceedings of the National Academy of Sciences},
volume = {119},
number = {12},
pages = {e2107151119},
year = {2022},
doi = {10.1073/pnas.2107151119},
URL = {https://www.pnas.org/doi/abs/10.1073/pnas.2107151119},
eprint = {https://www.pnas.org/doi/pdf/10.1073/pnas.2107151119},
}

@article{Fang2019,
    author = {Fang, Jianwen},
    title = "{A critical review of five machine learning-based algorithms for predicting protein stability changes upon mutation}",
    journal = {Briefings in Bioinformatics},
    volume = {21},
    number = {4},
    pages = {1285-1292},
    year = {2019},
    month = {07},
    issn = {1477-4054},
    doi = {10.1093/bib/bbz071},
    url = {https://doi.org/10.1093/bib/bbz071},
    eprint = {https://academic.oup.com/bib/article-pdf/21/4/1285/33584078/bbz071.pdf},
}

@article{Ying2019,
author = {Ying, Xue},
address = {Bristol},
copyright = {Published under licence by IOP Publishing Ltd},
issn = {1742-6588},
journal = {Journal of Physics: Conference Series},
language = {eng},
number = {2},
pages = {22022-},
publisher = {IOP Publishing},
title = {An Overview of Overfitting and its Solutions},
volume = {1168},
year = {2019},
}

@article{Dastile2020,
title = {Statistical and machine learning models in credit scoring: A systematic literature survey},
journal = {Applied Soft Computing},
volume = {91},
pages = {106263},
year = {2020},
issn = {1568-4946},
doi = {https://doi.org/10.1016/j.asoc.2020.106263},
url = {https://www.sciencedirect.com/science/article/pii/S1568494620302039},
author = {Xolani Dastile and Turgay Celik and Moshe Potsane}
}

@article{PredictionUncertainty,
author = {Lakshminarayanan, Balaji and Pritzel, Alexander and Blundell, Charles},
copyright = {http://arxiv.org/licenses/nonexclusive-distrib/1.0},
language = {eng},
title = {Simple and Scalable Predictive Uncertainty Estimation using Deep Ensembles},
year = {2016},
}

@misc{Ovadia2019,
  doi = {10.48550/ARXIV.1906.02530},
  
  url = {https://arxiv.org/abs/1906.02530},
  
  author = {Ovadia, Yaniv and Fertig, Emily and Ren, Jie and Nado, Zachary and Sculley, D and Nowozin, Sebastian and Dillon, Joshua V. and Lakshminarayanan, Balaji and Snoek, Jasper},
  
  title = {Can You Trust Your Model's Uncertainty? Evaluating Predictive Uncertainty Under Dataset Shift},
  
  publisher = {arXiv},
  
  year = {2019},
  
  copyright = {arXiv.org perpetual, non-exclusive license}
}

@article{Tohme2022,
  title={Reliable neural networks for regression uncertainty estimation},
  author={Tohme, Tony and Vanslette, Kevin and Youcef-Toumi, Kamal},
  journal={Reliability Engineering \& System Safety},
  pages={108811},
  year={2022},
  publisher={Elsevier}
}

@InProceedings{Klass2018,
author="Kl{\"a}s, Michael and Vollmer, Anna Maria",
editor="Gallina, Barbara
and Skavhaug, Amund
and Schoitsch, Erwin
and Bitsch, Friedemann",
title="Uncertainty in Machine Learning Applications: A Practice-Driven Classification of Uncertainty",
booktitle="Computer Safety, Reliability, and Security",
year="2018",
publisher="Springer International Publishing",
address="Cham",
pages="431--438",
isbn="978-3-319-99229-7"
}

@article{Bolker2009,
title = {Generalized linear mixed models: a practical guide for ecology and evolution},
journal = {Trends in Ecology \& Evolution},
volume = {24},
number = {3},
pages = {127-135},
year = {2009},
issn = {0169-5347},
doi = {https://doi.org/10.1016/j.tree.2008.10.008},
url = {https://www.sciencedirect.com/science/article/pii/S0169534709000196},
author = {Benjamin M. Bolker and Mollie E. Brooks and Connie J. Clark and Shane W. Geange and John R. Poulsen and M. Henry H. Stevens and Jada-Simone S. White}
}

@article{Pryseley2011,
title = {Estimating negative variance components from Gaussian and non-Gaussian data: A mixed models approach},
journal = {Computational Statistics \& Data Analysis},
volume = {55},
number = {2},
pages = {1071-1085},
year = {2011},
issn = {0167-9473},
doi = {https://doi.org/10.1016/j.csda.2010.09.002},
url = {https://www.sciencedirect.com/science/article/pii/S0167947310003452},
author = {Assam Pryseley and Clotaire Tchonlafi and Geert Verbeke and Geert Molenberghs}
}

@article{Oliveira2017,
issn = {0266-4763},
journal = {Journal of Applied Statistics},
pages = {1047--1063},
volume = {44},
publisher = {Sheffield City Polytechnic},
number = {6},
year = {2017},
title = {Negative variance components for non-negative hierarchical data with correlation, over-, and/or underdispersion},
language = {eng},
author = {Oliveira, IRC and Molenberghs, Geert and Verbeke, Geert and Demetrio, CGB and Dias, CTS},
}

@article{Zhu2021,
author = {Zhu, Rui and W{\"u}thrich, Mario V.},
address = {Cambridge, UK},
copyright = {The Author(s), 2020. Published by Cambridge University Press on behalf of Institute and Faculty of Actuaries},
issn = {1748-4995},
journal = {Annals of actuarial science},
language = {eng},
number = {2},
pages = {276-290},
publisher = {Cambridge University Press},
title = {Clustering driving styles via image processing},
volume = {15},
year = {2021},
}

@article{Gao2018,
author = {Gao, Guangyuan and W{\"u}thrich, Mario V.},
address = {Berlin/Heidelberg},
copyright = {EAJ Association 2018},
issn = {2190-9733},
journal = {European Actuarial Journal},
language = {eng},
number = {2},
pages = {383-406},
publisher = {Springer Berlin Heidelberg},
title = {Feature extraction from telematics car driving heatmaps},
volume = {8},
year = {2018},
}

@book{Fitzmaurice2014,
title = {Longitudinal data analysis: a handbook of modern statistical methods},
  author={Fitzmaurice, G. and Davidian, M. and Verbeke, G. and Molenberghs, G.},
  isbn={9781420011579},
  lccn={2008020681},
  series={Chapman \& Hall/CRC Handbooks of Modern Statistical Methods},
  url={https://books.google.be/books?id=zVBjCvQCoGQC},
  year={2008},
  publisher={CRC Press}
}

@article{MaidR,
title = {When stakes are high: Balancing accuracy and transparency with Model-Agnostic Interpretable Data-driven suRRogates},
journal = {Expert Systems with Applications},
volume = {202},
pages = {117230},
year = {2022},
issn = {0957-4174},
doi = {https://doi.org/10.1016/j.eswa.2022.117230},
url = {https://www.sciencedirect.com/science/article/pii/S0957417422006042},
author = {Roel Henckaerts and Katrien Antonio and Marie-Pier {C{\^o}t{\'e}}}
}

@article{Harrison2018,
author = {Harrison, Xavier A. and Donaldson, Lynda and Correa-Cano, Maria Eugenia and Evans, Julian and Fisher, David N. and Goodwin, Cecily E.D. and Robinson, Beth S. and Hodgson, David J. and Inger, Richard},
address = {LONDON},
copyright = {Copyright 2018 Elsevier B.V., All rights reserved.},
issn = {2167-8359},
journal = {PeerJ (San Francisco, CA)},
language = {eng},
number = {5},
pages = {e4794-e4794},
publisher = {Peerj Inc},
title = {A brief introduction to mixed effects modelling and multi-model inference in ecology},
volume = {2018},
year = {2018},
}

@book{Zuur2009,
  title={Mixed Effects Models and Extensions in Ecology with R},
  author={Zuur, A. and Ieno, E.N. and Walker, N. and Saveliev, A.A. and Smith, G.M.},
  isbn={9780387874586},
  lccn={2008942429},
  series={Statistics for Biology and Health},
  url={https://books.google.be/books?id=vQUNprFZKHsC},
  year={2009},
  publisher={Springer New York}
}
\clearpage

\bookmarksetup{startatroot}
\appendix
\section*{Appendices}
\addcontentsline{toc}{section}{Appendices}
\renewcommand{\thesection}{\Alph{section}}
\renewcommand{\thesubsection}{\Alph{section}.\arabic{subsection}}

\section{Jewell's hierarchical model: variance estimators}\label{App:JewellVariance}
In our analysis, we make use of the estimators proposed by \citet{OhlssonJewell}
\begin{equation}
	\begin{aligned}
		\sigma^2 &= \frac{1}{\sum_j \sum_k (T_{jk} - 1)} \sum_{i, j, k, t} w_{ijkt} (Y_{ijkt} - \bar{Y}_{\cdot jk \cdot})^2,\\
		\vspace{2mm}
		\sigma_{B}^2 &= \frac{\sum_j \sum_k w_{\cdot jk \cdot} (\bar{Y}_{\cdot jk \cdot} - \bar{Y}_{\cdot j \cdot \cdot})^2 - \hat{\sigma}^2 \sum_j (K_j - 1)}{w_{\cdot \cdot \cdot \cdot} - \sum_j \frac{\sum_k w_{\cdot jk \cdot}^2}{w_{\cdot j \cdot \cdot}} },\\
		\vspace{2mm}
		\sigma_{I}^2 &= \frac{\sum_j z_{j \cdot} (\bar{Y}_{\cdot j \cdot \cdot}^z - \bar{Y}_{\cdot \cdot \cdot \cdot}^z)^2 - \hat{\sigma_B}^2 (J - 1)}{z_{\cdot \cdot} - \frac{\sum_j z_{j \cdot}^2}{z_{\cdot \cdot}}},
	\end{aligned}
\end{equation}
\noindent
where 
\begin{equation}\label{eq:JewellRE}
	\begin{aligned}
		\bar{Y}_{\cdot j \cdot \cdot} &= \frac{\sum_k w_{\cdot jk \cdot} \bar{Y}_{\cdot jk \cdot}}{\sum_k w_{\cdot jk \cdot}} \ \ \ \text{and} \ \ \ 
		\bar{Y}_{\cdot \cdot \cdot \cdot}^z &= \frac{\sum_j z_{j \cdot} \bar{Y}_{\cdot j \cdot \cdot}^z}{\sum_j z_{j \cdot}}.
	\end{aligned}
\end{equation}
In the above equations, $T_{jk}$ denotes the number of observations in group $(j, k)$, $K_j$ the number of branches in industry $j$ and $J$ the number of industries. 

To estimate $\mu$, we use
\begin{equation}
	\begin{aligned}
		\hat{\mu} = \frac{\sum_j q_j \bar{Y}^z_{\cdot j \cdot \cdot}}{\sum_j q_j}.
	\end{aligned}
\end{equation}
\noindent

\clearpage

\section{Random effect estimates}\label{App:REs}
\begin{figure}[H]
	\centering
	\caption{\label{fig:LMMREGroup1}LMM: Random effects estimates of the branches within industries}
	\makebox[\textwidth][c]{\includegraphics[width = \textwidth]{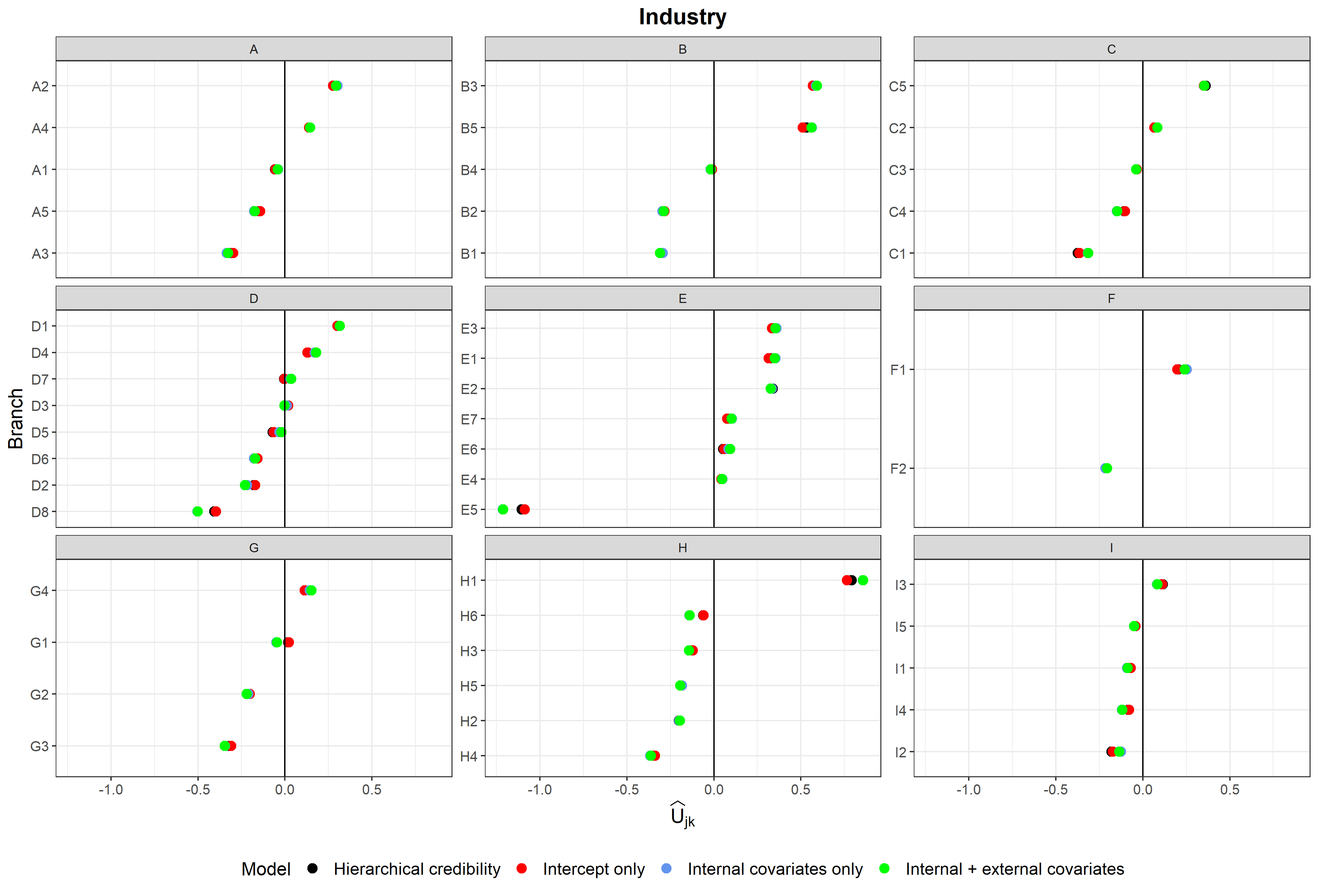}}
\end{figure}

\begin{figure}[H]
	\centering
	\caption{\label{fig:LMMREGroup2}LMM: Random effects estimates of the branches within industries (\textit{continued})}
	\makebox[\textwidth][c]{\includegraphics[width = \textwidth]{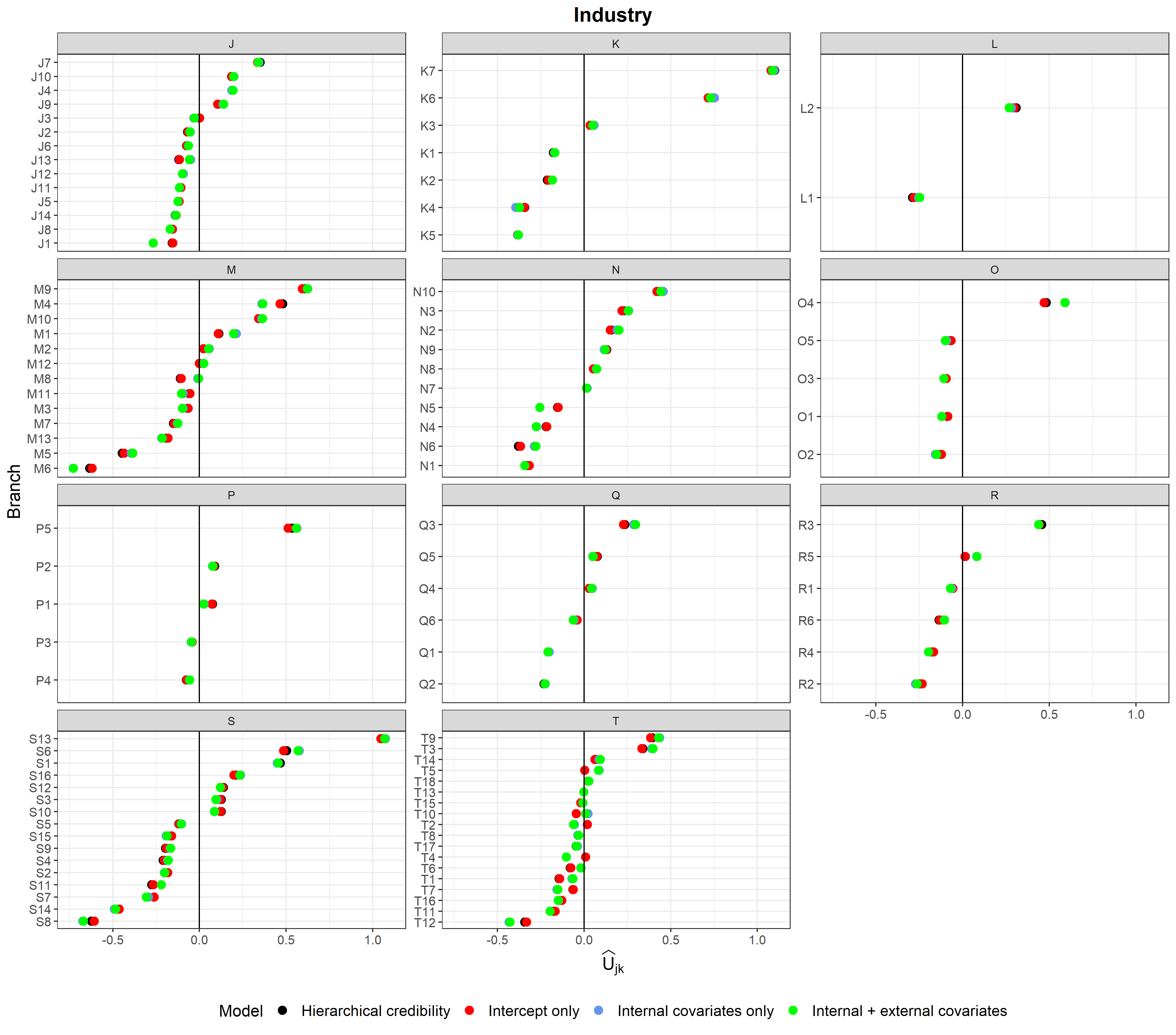}}
\end{figure}

\begin{figure}[H]
	\centering
	\caption{\label{fig:TweedieREGroup1}Tweedie GLMM: Random effects estimates of the branches within industries}
	\makebox[\textwidth][c]{\includegraphics[width = \textwidth]{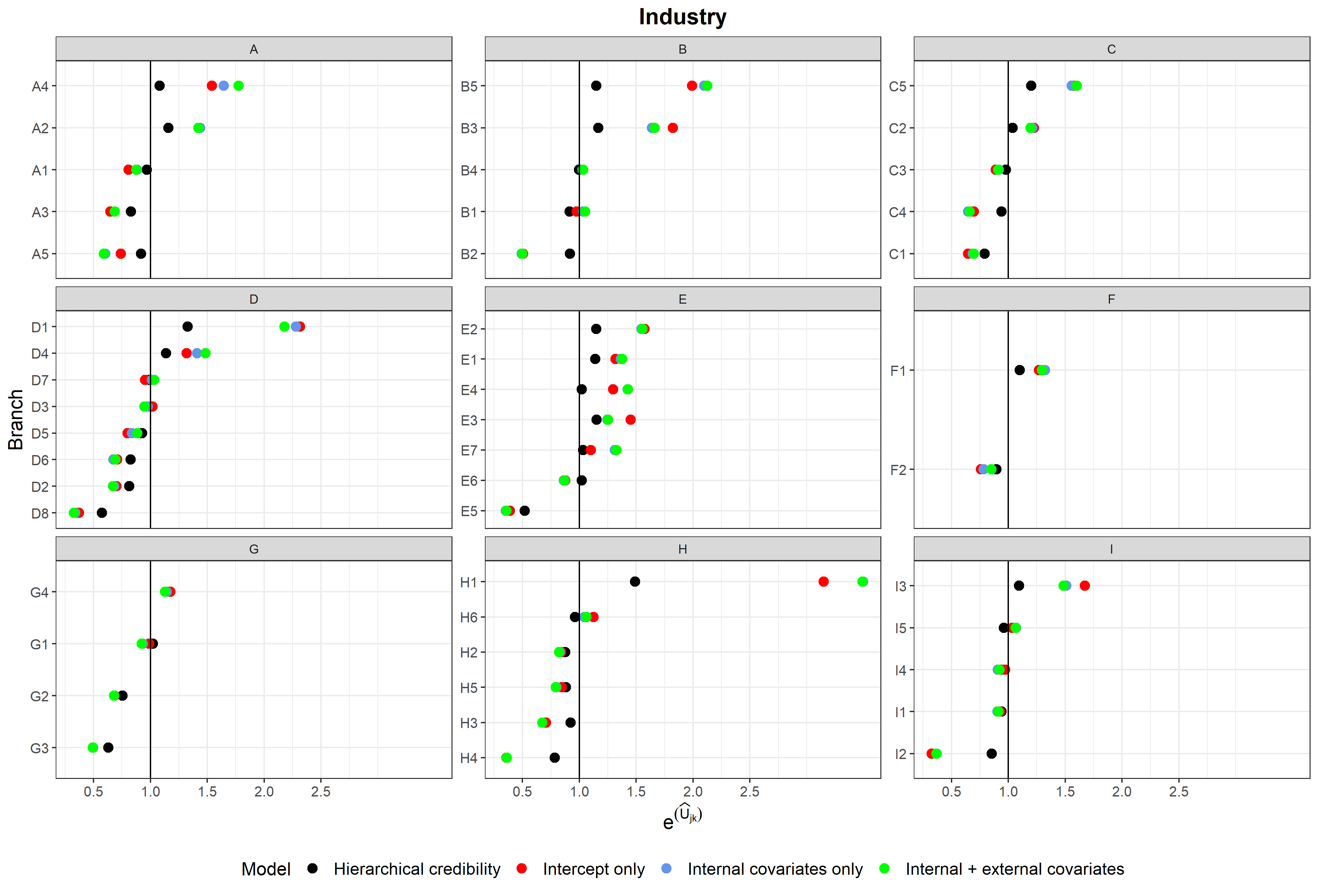}}
\end{figure}

\begin{figure}[H]
	\centering
	\caption{\label{fig:TweedieREGroup2}Tweedie GLMM: Random effects estimates of the branches within industries (\textit{continued})}
	\makebox[\textwidth][c]{\includegraphics[width = \textwidth]{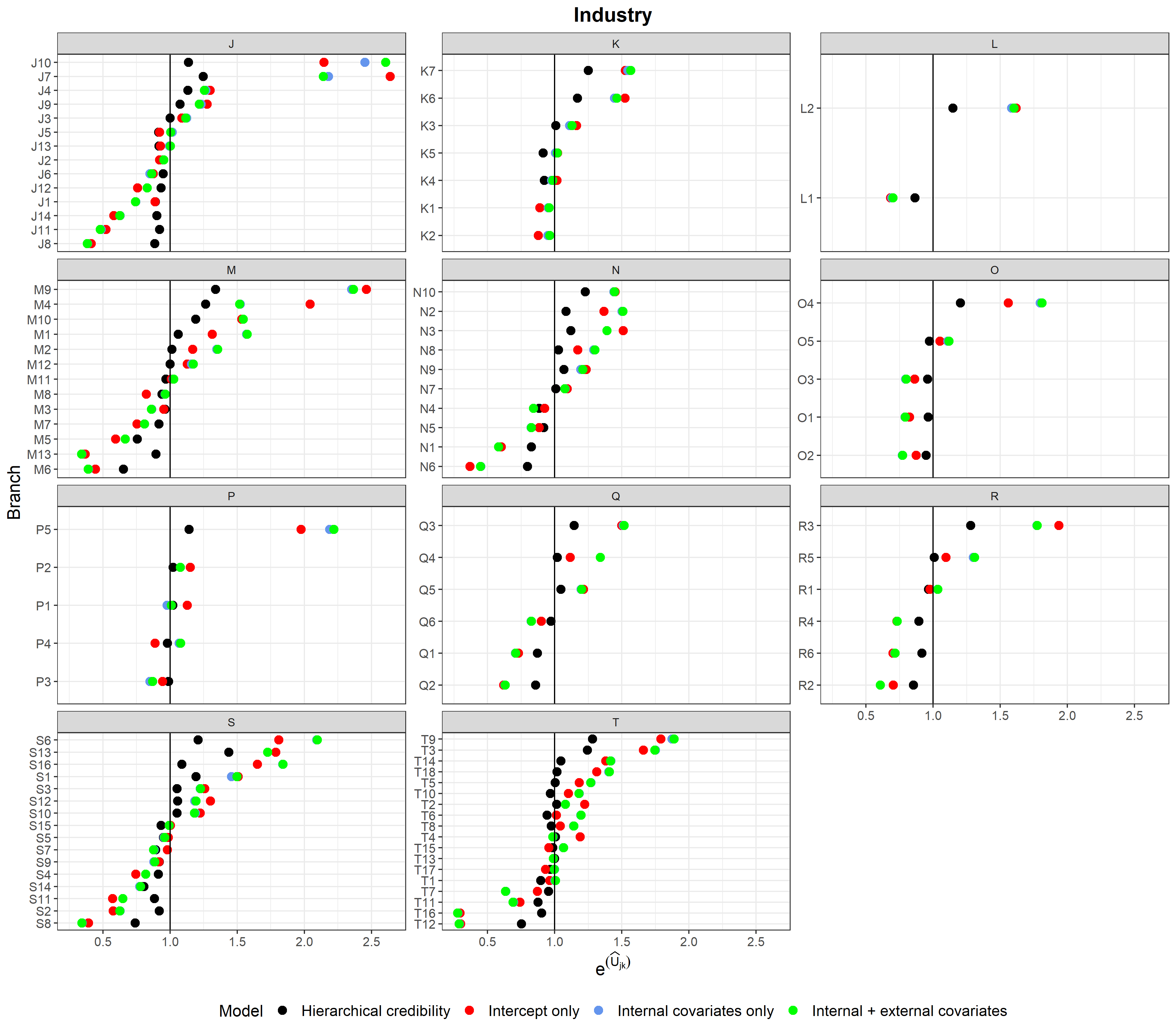}}
\end{figure}
\clearpage

\end{document}